\documentclass[12pt]{article}

\usepackage[T1]{fontenc}
\usepackage{lmodern}

\usepackage{setspace}
\setdisplayskipstretch{1}

\usepackage[margin=1.25in]{geometry}

\usepackage[dvipsnames]{xcolor}
\usepackage{pdfpages}
\usepackage{float}
\usepackage{multirow}
\usepackage{caption}
\usepackage{subcaption}
\usepackage{epigraph}
\usepackage{titlesec}

\titleformat{\section}
  {\normalfont\large\bfseries}{\thesection}{1em}{}
\titleformat{\subsection}
  {\normalfont\normalsize\bfseries}{\thesubsection}{1em}{}

\usepackage{mathtools}
\usepackage{amssymb}
\usepackage{amsthm}
\usepackage{amsfonts}
\usepackage{aliascnt}
\usepackage{accents}
\usepackage{dutchcal}
\usepackage{bbm}

\usepackage{enumitem}
\usepackage{sgame}
\usepackage{tikz}
\usepackage{tikz-cd}
\usetikzlibrary{calc,shapes,arrows}
\usepackage{tcolorbox}
\usepackage{listings}
\usepackage[normalem]{ulem}

\usepackage[round]{natbib}
\DeclareMathOperator{\cl}{cl}

\newcommand{\E}{\mathbb{E}}
\newcommand{\R}{\mathbf{R}}

\newcommand{\D}{\mathcal{D}}

\newcommand{\K}{\mathcal K}
\newcommand{\V}{\mathcal V}
\newcommand{\A}{\mathcal A}
\newcommand{\M}{\mathcal M}
\DeclareMathOperator{\Eq}{\mathsf{Eq}}
\DeclareMathOperator{\dist}{dist}
\setlist[enumerate]{label=\textnormal{(\roman*)},itemsep=3pt,topsep=5pt}

\newtheorem{theorem}{Theorem}[section]

\newaliascnt{proposition}{theorem}
\newtheorem{proposition}[proposition]{Proposition}
\aliascntresetthe{proposition}

\newaliascnt{lemma}{theorem}
\newtheorem{lemma}[lemma]{Lemma}
\aliascntresetthe{lemma}

\newaliascnt{corollary}{theorem}
\newtheorem{corollary}[corollary]{Corollary}
\aliascntresetthe{corollary}

\newaliascnt{claim}{theorem}
\newtheorem{claim}[claim]{Claim}
\aliascntresetthe{claim}

\theoremstyle{definition}

\newaliascnt{definition}{theorem}
\newtheorem{definition}[definition]{Definition}
\aliascntresetthe{definition}

\newaliascnt{example}{theorem}

\aliascntresetthe{example}

\newaliascnt{assumption}{theorem}

\aliascntresetthe{assumption}

\newaliascnt{condition}{theorem}

\aliascntresetthe{condition}

\newaliascnt{question}{theorem}

\aliascntresetthe{question}

\newaliascnt{remark}{theorem}

\aliascntresetthe{remark}

\newaliascnt{remarks}{theorem}

\aliascntresetthe{remarks}

\newaliascnt{aside}{theorem}

\aliascntresetthe{aside}

\newaliascnt{note}{theorem}

\aliascntresetthe{note}

\usepackage{thmtools}
\usepackage{thm-restate}

\usepackage{hyperref}

\hypersetup{
    colorlinks=true,
    linkcolor=OrangeRed,
    filecolor=Thistle,
    urlcolor=Thistle,
    citecolor=Thistle,
}

\usepackage[nameinlink]{cleveref}

\crefname{theorem}{theorem}{theorems}
\Crefname{theorem}{Theorem}{Theorems}
\crefname{proposition}{proposition}{propositions}
\Crefname{proposition}{Proposition}{Propositions}
\crefname{lemma}{lemma}{lemmas}
\Crefname{lemma}{Lemma}{Lemmas}
\crefname{corollary}{corollary}{corollaries}
\Crefname{corollary}{Corollary}{Corollaries}
\crefname{claim}{claim}{claims}
\Crefname{claim}{Claim}{Claims}
\crefname{definition}{definition}{definitions}
\Crefname{definition}{Definition}{Definitions}
\crefname{example}{example}{examples}
\Crefname{example}{Example}{Examples}
\crefname{assumption}{assumption}{assumptions}
\Crefname{assumption}{Assumption}{Assumptions}
\crefname{axiom}{axiom}{axioms}
\Crefname{axiom}{Axiom}{Axioms}

\makeatletter
\let\cref@old@isrefconsecutive\cref@isrefconsecutive
\def\cref@isrefconsecutive#1#2{%
  \begingroup
    \def\cref@assumptiontype{assumption}%
    \cref@gettype{#1}{\cref@typea}%
    \ifx\cref@typea\cref@assumptiontype
      \endgroup
      \@cref@refconsecutivefalse
    \else
      \endgroup
      \cref@old@isrefconsecutive{#1}{#2}%
    \fi
}
\makeatother

\crefname{condition}{condition}{conditions}
\Crefname{condition}{Condition}{Conditions}
\crefname{question}{question}{questions}
\Crefname{question}{Question}{Questions}
\crefname{remark}{remark}{remarks}
\Crefname{remark}{Remark}{Remarks}
\crefname{remarks}{remarks}{remarks}
\Crefname{remarks}{Remarks}{Remarks}
\crefname{aside}{aside}{asides}
\Crefname{aside}{Aside}{Asides}
\crefname{note}{note}{notes}
\Crefname{note}{Note}{Notes}
\crefname{appendix}{appendix}{appendices}
\Crefname{appendix}{Appendix}{Appendices}

\newcommand{\secref}[1]{\hyperref[#1]{\S\ref*{#1}}}
\newcommand{\conditionref}[1]{\hyperref[#1]{\textcolor{Blue}{\textnormal{\ref*{#1}}}}}
\newcommand{\Conditionref}[1]{\hyperref[#1]{\textcolor{Blue}{\textnormal{\ref*{#1}}}}}

\definecolor{backcolour}{rgb}{0.63,0.79,0.95}
\lstdefinestyle{mystyle}{
  backgroundcolor=\color{backcolour},
  basicstyle=\ttfamily\footnotesize,
  breakatwhitespace=false,
  breaklines=true,
  captionpos=b,
  keepspaces=true,
  numbers=left,
  numbersep=5pt,
  showspaces=false,
  showstringspaces=false,
  showtabs=false,
  tabsize=2
}
\begin{document}
\title{Principal Revelations of the Revelation Principle}
\author{Mark Whitmeyer\thanks{For KS.}}
\date{\today}
\maketitle
\begin{abstract}
I prove a converse to the revelation principle, arguing that it is a principle of optimism: if it must hold for all environments (in fact, even just those with private values and independent types) the principal must assume her favorite equilibrium is played by the agents. When the principal knows only that the prior lies in some set, she must additionally assume it is her favorite one. With private values, independent types, and quasilinear payoffs, however, unrestricted transfers allow the principal to approximate any mechanism's equilibrium payoff via a direct mechanism whose unique equilibrium is truthful.
\end{abstract}

A principal designs a game, \textit{viz.,} a mechanism, for a group of agents to play. The revelation principle says that this problem of game design can be simplified dramatically: it is without loss of optimality for the principal to replace any (simultaneous-move) game for the agents that she could think of with a simple game in which the agents merely ``report'' their private information, with the new mechanism itself carrying out any of the complicated behavior the principal might desire. It is the principle that any game can be replaced with another (much) simpler one, without harming the principal.

But the principal has replaced one game with another, and different games have different equilibria. What does the revelation principle holding reveal about how the principal evaluates games–mechanisms–and the equilibria played therein?

The thesis of this paper is that the revelation principle is \textbf{a principle of optimism}: if it holds universally (in a sense I make (more) precise a few paragraphs down), the principal must accommodate equilibrium multiplicity by assuming her most preferred one is played; and, when she does not know the exact (random) features of the environment, the principal must evaluate her mechanisms assuming the best (for her) distribution.

My setting is basic. A principal and agents share a common full-support prior over agents' types. A principal designs (and commits to) a \textit{finite mechanism} $M$, which specifies a finite nonempty set of messages $A_i$ for each agent and an outcome rule $g\colon A\to\Delta(X)$ (for a nonempty set of outcomes \(X\)). Each agent observes the mechanism and his own type before simultaneously sending a message. Any such mechanism corresponds to a game of incomplete information played by the agents, which has a corresponding set of equilibria, each of which delivers to the principal a numerical payoff. The principal evaluates a mechanism according to a function $\V\colon\K(\R)\to\R$ whose domain is the collection of nonempty compact subsets of $\R$, which has as its argument the set of all equilibrium payoffs for the principal, and which I assume is continuous in the Hausdorff distance and produces an output lying between the worst and best equilibrium payoffs (is \textit{internal} if one wants a term).

I first prove that it is without loss of optimality in every environment for the principal to replace any mechanism with a truthful direct mechanism--that is to say, the revelation principle holds universally--if and only if \(\V(S) = \max S\) for every \(S \in \K(\R)\), i.e., the principal must evaluate any payoff multiplicity in the most optimistic manner and assume that the best equilibrium will be played.\footnote{"If" is of course, classical. My contribution is the converse.} Moreover, this equivalence holds even when environments are restricted to those with private values and independent types.

In my second result, I assume that the principal only knows the agents' common prior to lie in some nonempty compact set \(\mathcal{P}\). Now, the principal evaluates a mechanism by first, for every prior in \(\mathcal{P}\), assigning a payoff to the set of equilibrium payoffs according to \(\V\) (as above). She then takes the closure over all priors in \(\mathcal{P}\) of the set of all such payoffs and evaluates such sets via another internal, Hausdorff-continuous function.

It is surprisingly demanding for the revelation principle to hold universally with this additional layer of ambiguity. First, I show that it necessitates a condition I term \textit{common interim beliefs}, which, for environments with at least two agents with at least two types, requires that the principal \textbf{knows the agents'  common prior}. Having assumed common interim beliefs, optimism emerges even more strongly: both layers of evaluation must be \(\max\).

I finish the paper by introducing transfers, revealing that in environments in which the principal's and agents' payoffs are additively separable therein, the principal's evaluation criterion is approximately irrelevant. Namely, for every principal payoff from a Bayesian Nash equilibrium induced by some mechanism, there is a direct mechanism whose unique equilibrium is truthful and under which the principal's equilibrium payoff is arbitrarily close to the original one. The key insight is that the transfers in any strict Bayesian incentive compatible direct mechanism can be modified to yield the allocation unchanged, expected transfers under truth telling unchanged, but with truth telling as the unique equilibrium of the new game.

\subsection*{Related papers}

The standard revelation principle is that an equilibrium outcome of an arbitrary mechanism can be reproduced as the truthful equilibrium outcome of a direct mechanism.\footnote{See \citet{DHM1979} and \citet{Myerson1982}.} The original mechanism's other equilibria need not be reproduced, nor does the principle discipline additional equilibria of the direct mechanism. Full implementation instead requires that the full set of equilibrium outcomes of the mechanism's game be the desired one.\footnote{\citet{PalfreySrivastava1989} proffer an example of an allocation rule that an indirect mechanism fully implements in spite of every equivalent direct mechanism having additional equilibrium outcomes. See also \citet{Repullo1985}, \citet{PalfreySrivastava1989b}, \citet{MR1990}, \citet{Jackson1991}, and \citet{SV2010}.}

Closer yet are the papers that prescribe or study a designer's treatment of equilibrium multiplicity. There are a number of works in the information design literature--some impose selection of the designer's favorite equilibrium; others, adversarial selection of her least favorite one.\footnote{\citet{BergemannMorris2016} and \citet{Taneva2019} study designer-optimal information design in games. \citet{BM2019} discuss why adversarial equilibrium selection breaks the usual directification argument. \citet{MPT2020} allow for different selection rules and study an application under adversarial selection; see also \citet{MorrisOyamaTakahashi2024}.} \citet{BMV2024} endow their designer with an arbitrary Hausdorff-continuous objective over the set of equilibrium outcomes and shows that simple information structures suffice for a broad class of information-design problems. Informationally robust mechanism design prescribes that a mechanism be evaluated by its worst performance across information structures and equilibria.\footnote{E.g., \citet{BergemannBrooksMorris2016}, \citet{BrooksDu2021}, and \citet{BrooksDu2024}.}

There has also been a significant endeavor to weaken the common-knowledge assumptions on agents' beliefs. In turn, prior-independent mechanism design seeks mechanisms that perform well across a class of distributions. In particular environments, indirect mechanisms can outperform every truthful prior-independent mechanism--the loss measured by the \textit{revelation gap}.\footnote{\citet{BM2005} study mechanism design on richer type spaces. Also glance at \citet{ChungEly2007} and \citet{BergemannMorris2009}. For prior-independent mechanism design, see, e.g., \citet{DhangwatnotaiRoughgardenYan2015}. \citet{FengHartline2018} introduce the revelation gap, while \citet{FengHartlineLi2021} establish a nontrivial revelation gap for revenue maximization.} There is also earlier work in which an indirect mechanism uses bidders' knowledge of distributions unknown to the seller.\footnote{\citet{CaillaudRobert2005}. or \citet{ChoiKim1999} in a public-good setting.} My agents know the prior, while the principal must choose one mechanism for all priors in a given set.

Finally, my result with unrestricted transfers in \secref{sec:unrestricted-transfers} also relates to work on eliminating undesirable equilibria. In auctions with independent types, \citet{dAspremontCremerGerardVaret2005} enlarge bidders' message spaces so that every equilibrium delivers the seller the revenue from a selected equilibrium. Other papers instead use transfers to control equilibrium multiplicity in direct mechanisms.\footnote{\citet{Mathevet2010} and \citet{MathevetTaneva2013} provide conditions for uniqueness in supermodular mechanisms. \citet{OllarPenta2017,OllarPenta2023} study full implementation under restrictions on agents' beliefs, allowing interdependent valuations and weaker common-knowledge assumptions.} I argue that a principal can use transfers to eliminate inferior equilibria rather than assume they will not be played. In a companion paper to this one, \citet{obedprinc2026}, I characterize when the obedience principle holds universally in games with and without incomplete information. In the former, like here, the principal's evaluation must be \(\max\); without, it is a broader class that includes both \(\max\) and \(\min\).

\section{Sketches}\label{sec:nec}

As I noted in the introduction, it is the necessity direction that is the contribution of my two theorems. For a known prior, the intuition for why optimism must hold is simple. Suppose for the sake of contraposition that the principal's evaluation \(\V\) does not always select the largest equilibrium payoff from every nonempty compact set, but is continuous in the Hausdorff distance.\footnote{I haven't introduced any of this formally yet and do so in the next section. It should nevertheless be easy to follow along.} This continuity then guarantees that I can find a finite set of equilibrium payoffs \(F = \{b,a_1,\ldots,a_k\}\), with \(b = \max{F} > \V(F)\).

For simplicity, assume first that \(F=\{b,a\}\) and take an environment with two agents, each of whom has an independent uniform binary type (in \(\{0,1\}\)), and nine outcomes \(X=(\{0,1\}\cup\{C\})^2\) where \(C\) is an additional symbol. Take also the direct mechanism \(D^0\) that asks each agent to report either \(0\) or \(1\) and, after the agents choose these reports \((r_1,r_2)\), selects the outcome \(x=(r_1,r_2)\). I choose the agents' payoffs so that each wants to use the same reporting rule as the other. More precisely, each agent receives payoff \(0\) when both agents report truthfully or lie, and payoff \(-1\) when one reports truthfully and the other lies.

As a result, the agents face a \textit{coordination game} with two strict (Bayes-Nash) equilibria: both lie, or both tell the truth. There is also a mixed-strategy equilibrium in which each type of each agent tells the truth and lies with equal probability. In fact, every remaining equilibrium induces truth telling with probability \(1/2\) for each agent, after averaging across his types.

\begin{table}[ht]
\centering
\begingroup
\small
\setlength{\arraycolsep}{5pt}
\renewcommand{\arraystretch}{1.3}

\[\begin{array}{c@{\qquad}c} \begin{array}{c} (t_1,t_2)=(0,0) \\[0.3em] \begin{array}{c|cc} & r_2=0 & r_2=1 \\ \hline r_1=0 & (0,0\,;\,b) & (-1,-1\,;\,\tilde a) \\ r_1=1 & (-1,-1\,;\,\tilde a) & (0,0\,;\,a) \end{array} \end{array} & \begin{array}{c} (t_1,t_2)=(0,1) \\[0.3em] \begin{array}{c|cc} & r_2=0 & r_2=1 \\ \hline r_1=0 & (-1,-1\,;\,\tilde a) & (0,0\,;\,b) \\ r_1=1 & (0,0\,;\,a) & (-1,-1\,;\,\tilde a) \end{array} \end{array} \\[4em] \begin{array}{c} (t_1,t_2)=(1,0) \\[0.3em] \begin{array}{c|cc} & r_2=0 & r_2=1 \\ \hline r_1=0 & (-1,-1\,;\,\tilde a) & (0,0\,;\,a) \\ r_1=1 & (0,0\,;\,b) & (-1,-1\,;\,\tilde a) \end{array} \end{array} & \begin{array}{c} (t_1,t_2)=(1,1) \\[0.3em] \begin{array}{c|cc} & r_2=0 & r_2=1 \\ \hline r_1=0 & (0,0\,;\,a) & (-1,-1\,;\,\tilde a) \\ r_1=1 & (-1,-1\,;\,\tilde a) & (0,0\,;\,b) \end{array} \end{array} \end{array}\]

\endgroup
\caption{The reporting game induced by \(D^0\), where \(\tilde a=(3a-b)/2\). Rows are agent \(1\)'s reports, columns are agent \(2\)'s reports, with the convention \((u_1,u_2\,;\,v)\).}
\label{tab:binary-state-games}
\end{table}

I make the principal's payoff \(b\) when both agents report truthfully and \(a\) when both lie. When one agent tells the truth and the other lies, I give the principal \((3a-b)/2\). At every outcome involving \(C\), I give the principal \(a\). Her expected payoff in any remaining equilibrium is then \(\frac14 b+\frac14 a+\frac12\frac{3a-b}{2}=a\), and so she receives \(b\) in the truthful equilibrium and \(a\) in every other equilibrium. By construction, the equilibrium-payoff set of \(D^0\) is exactly \(\{b,a\}\).

Returning to the general finite \(F\), I use the same idea with \(k+1\) types and \(k+1\) available reports for each of the two agents, and \(D^0\) again selects the pair of reports as the outcome (where the set of outcomes is now \(X=(\{0,\ldots,k\}\cup\{C\})^2\)). After observing his type, an agent chooses how far ``ahead'' of that type to report, with the labels understood modulo \(k+1\)--think of the types arranged around a circle. To generate the analogous coordination motives, I choose the agents' payoffs so that each wants his report to be shifted from his type by the same amount as the other agent's report is shifted from his.

Consequently, for every \(j\), both agents using shift \(j\) is a strict equilibrium. Shifting by \(0\) is truth telling, and I give the principal payoff \(a_j\) when both agents use shift \(j\) (so, \(b\) when they tell the truth, \textit{viz.}, \(b = a_0\)) and \(a_k\) at every outcome involving \(C\). The quadratic-loss payoff I endow the agents with makes it so that, in every remaining equilibrium, the distribution of each agent's shift assigns probability one half to each of two consecutive shifts. As I did for the two-payoff \(F\), I choose the principal's payoff when the agents use different shifts so that each such mixed equilibrium gives her one of the same values \(a_0,\ldots,a_k\). Ultimately, this makes it so that the equilibrium-payoff set of \(D^0\) is exactly \(F\).

\begin{figure}[ht]
\centering
\begin{tikzpicture}[
    scale=1.1,
    >=stealth,
    type/.style={
        circle,
        draw,
        fill=white,
        minimum size=8mm,
        inner sep=0pt
    },
    shiftarrow/.style={
        ->,
        thick,
        shorten <=2pt,
        shorten >=2pt
    }
]
\def\r{1.55}

\foreach \i in {0,...,4} {
    \node[type] (t\i) at ({90-72*\i}:\r) {$\i$};
}

\foreach \i/\j in {0/1,1/2,2/3,3/4,4/0} {
    \draw[shiftarrow]
        (t\i) to[bend left=18] (t\j);
}

\node at (0,0) {\(\text{Shift by }1\)};
\end{tikzpicture}
\caption{Shifting by \(1\) with five types: each type reports the next label clockwise. More generally, shift \(j\) moves each type \(j\) positions clockwise, modulo \(k+1\).}
\label{fig:reporting-shift}
\end{figure}
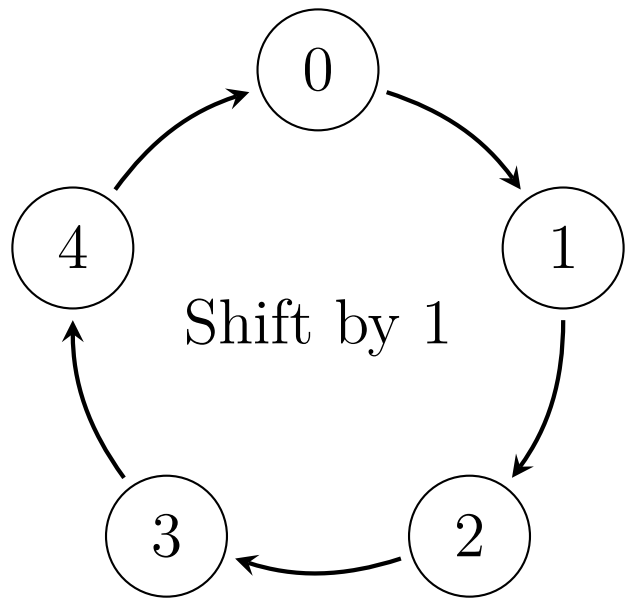

With this construction in hand, I then take an indirect mechanism \(M^+\) that makes \(C\) available as an additional message to each agent and selects the message profile as the outcome. I choose the agents' payoffs at outcomes involving \(C\) so that reporting \(C\) strictly dominates every false report and, once the false reports have been eliminated, truthful reporting strictly dominates reporting \(C\). In short, this indirect mechanism is such that the induced game is dominance solvable, with a unique equilibrium in which the agents report truthfully and the principal receives her maximal payoff \(b\).

Why can no direct mechanism also obtain \(b\) while eliminating the lower-payoff equilibria? First, recall that \(D^0\) does not produce an outcome involving \(C\). Of course, a direct mechanism could do so, but because the prior has full support, any equilibrium giving the principal \(b\) must produce the type profile as the outcome at every type profile. Moreover, this exact matching of type profiles with outcome profiles means that different types of an agent must use disjoint sets of reports and, since a direct mechanism gives each agent exactly as many reports as types, each type must use one report, and the equilibrium reporting rule must be a relabeling of the types. As a result, the mechanism cannot produce an outcome involving \(C\) after any report profile. This pins down the entire direct mechanism, which up to relabeling is \(D^0\), with the already established \(F\) as the set of equilibrium payoffs. Finally, I rule out direct mechanisms that come arbitrarily close to giving the principal \(b\) by appealing to the strictness of equilibria, compactness of the space of direct outcome rules and strategy profiles, and the continuity of \(\V\).

\subsection{Unknown prior}

Specializing to singleton sets of priors, the first theorem immediately pins down the necessary form of \(\V\). Now, however, ambiguity over the prior brings with it two further issues. First, the direct mechanism constructed from an equilibrium under one prior need not have a truthful equilibrium under another. An agent evaluates a deviation by averaging its payoff consequences using his belief about the other agents' types conditional on his own type. If every prior in the set of priors entertained by the principal, \(\mathcal P\), induces the same such conditional beliefs (which I term \textit{common interim beliefs}), the truth-telling inequalities are identical under every prior.

Common interim beliefs is necessary for the revelation principle: if it does not hold, I can choose payoffs so that the same type prefers truth telling under one prior and deviating under another. I also make the principal receive her maximal payoff when and only when the outcome equals the type profile. Any direct mechanism attaining that payoff is then pinned down by this requirement and cannot avoid the conflicting truth-telling inequalities. And so, no direct mechanism can then reproduce the desired outcome while inducing truth-telling under every prior, regardless of how the principal evaluates the resulting payoffs.

Suppose now that \(\mathcal P\) has common interim beliefs. To identify the restriction on \(\A\), fix \(a<b\), and take an environment with two agents, the first with only one possible type and the second with two. There are two possible priors \(p^0\) and \(p^1\), the former making type \(0\) more likely and the latter  type \(1\). There are also two outcomes, \(0\) and \(1\), and I assign agent \(1\) and the principal a high payoff \(H\) when the outcome matches agent \(2\)'s type and a low payoff \(L\) otherwise. Moreover, I choose \(H\) and \(L\) so that outcome \(0\) gives each of them expected payoff \(b\) under \(p^0\) and \(a\) under \(p^1\), while outcome \(1\) gives them \(a\) under \(p^0\) and \(b\) under \(p^1\). I make both types of agent \(2\) strictly prefer outcome \(0\): each receives utility \(1\) from outcome \(0\) and \(0\) from outcome \(1\).

Now consider the following indirect mechanism \(M^*\): agent 1 is the dictator and may choose either \(0\) or \(1\). Consequently, he chooses \(0\) under \(p^0\) and \(1\) under \(p^1\). As a result, the principal receives \(b\) no matter the prior. Now consider instead any direct mechanism in which agents report truthfully under both priors. This cannot condition on agent 2's information (his knowledge of his type), because both of his types strictly prefer outcome \(0\). But agent \(1\)'s report is useless--he has no private information \textbf{other than the prior}, but that is outside of the mechanism.\footnote{If the agents' types are augmented to include the prior as well, the revelation principle is again equivalent to optimism over the equilibrium played by the agents. See \secref{subsec:full-information}.} As a result, even though we are selecting the best equilibrium for the principal for both \(p\)s,  if the set of prior-specific values for the principal includes \(b\), it must be \(\left\{a,b\right\}\), and consequently, appealing to the continuity (and internality) of \(\mathcal{A}\), it must be that \(\A(\{a,b\})=b\). Finally, this logic extends easily to finite \(F\)s, with continuity of \(\A\) taking it further to compact sets.

\subsection{Transfers make the evaluation criterion moot}

Perhaps surprisingly, when the principal can choose monetary transfers freely and utilities (of principal and agent) are additively separable in transfers, the evaluation criterion becomes irrelevant in the sense that there is a truthful direct mechanism that allows the principal to approximate her expected payoff in the unique equilibrium of the game.

The key idea is that I can modify the agents' payments in any strictly Bayesian incentive-compatible (BIC) direct mechanism so that each agent's expected payment from any report when the other agents report truthfully is unchanged but may be different when the other agents do not report truthfully in such a way that every other equilibrium is eliminated. The transfer transformation works as follows. I start with the strictly BIC direct mechanism and, for each agent (assuming that multiple agents have more than one type or else I'm done), I choose one report to serve as a reference and pair every other report with a report of another agent, using each report at most once. If any reports remain unpaired, they all belong to one agent, whom I single out. If none do, I single out any agent with more than one type. For each such pair, I subtract from the indicator of each report the probability with which truth telling generates that report and multiply the two resulting terms, adding it to one agent's payment and subtracting it from the other's. This adjustment to payments has expected value zero for every report of one agent when his opponents report truthfully, leaving every comparison between truth telling and other reports unchanged, and so honesty remains a strict equilibrium. Before I do the scaling below, these adjustments are also budget-neutral for every report profile.

To see what these adjustments accomplish away from truth telling, temporarily ignore the original mechanism and let the agents care only about the adjustments. Every agent can obtain zero by choosing each report with the \textit{ex ante} probability of the corresponding type. Consider any equilibrium of this adjustment-only game. Since each agent can obtain zero, every agent's equilibrium payoff is at least zero. Since the agents' payoffs sum to zero at every report profile, every agent's equilibrium payoff must equal zero. Against the opponents' equilibrium strategies, no report can give an agent more than zero, yet choosing reports according to the \textit{ex ante} (full-support) probability produces expected payoff of zero, so every report must yield expected payoff zero against the opponents' equilibrium strategies.

Now consider two reports that I paired in the construction. Comparing one agent's payoff from his paired report with his payoff from his reference report shows that the other paired report must be used with the \textit{ex ante} probability of the corresponding type, which is exactly the probability with which truth telling would generate that report. Making the same comparison for the other agent shows that the first paired report must also be used with the \textit{ex ante} probability of its corresponding type. Repeating these comparisons across all pairs shows that, in every equilibrium of this adjustment-only game, every agent except the one I singled out uses each report with the \textit{ex ante} probability of the corresponding type.

Returning to the game induced by the modified mechanism, I next scale the payment adjustments, making them grow only linearly for the singled-out agent and quadratically for the rest. Suppose for the sake of contradiction that a nontruthful equilibrium survives for arbitrarily large adjustment scales, allowing me to then choose a sequence of scales tending to infinity and a nontruthful equilibrium at each scale. The payoffs in the original mechanism are bounded. Moreover, for each agent and each possible report, as we're at equilibrium, no type of any agent has a profitable deviation. I average these inequalities over the agent's types and divide by the size of his payment adjustment. Because payoffs in the original mechanism are bounded, their contribution to incentives disappears as the scale grows. And so any limit of the agents' report distributions must then be an equilibrium of the game formed by the adjustments alone. So, along the purported sequence of equilibria, the probability with which every such agent uses each report converges to the \textit{ex ante} probability of the corresponding type.

Why must the remaining agent do the same? Once his opponents' report distributions are close to those generated by truth telling, strict Bayesian incentive compatibility makes every false report strictly worse than truth. His gain from switching fully to truth is thus bounded below by a positive constant times his probability of misreporting. On the other hand, I can weight and add the agents' equilibrium inequalities so that the payment adjustments cancel. Because the other agents' adjustments grow quadratically while his grows only linearly, the resulting inequality multiplies his gain by the common scale while leaving the other agents' gains unscaled. Those other gains remain bounded, so his gain, and hence his probability of misreporting, must converge to zero.

As a result, all agents' report distributions are close to those produced by truth telling. Strict Bayesian incentive compatibility now makes every agent's gain from switching fully to truth positive, strictly so whenever that agent misreports with strictly positive probability. Yet, if I divide each inequality stating that an agent cannot profitably switch to truth in the modified mechanism by the coefficient multiplying his payment adjustment and then add these inequalities, the adjustment terms cancel. The resulting weighted sum of the agents' gains under the original mechanism from switching to truth is weakly negative. Since every one of these gains is positive, every one must equal zero, so no agent misreports, contradicting the supposed existence of another equilibrium.

\section{The Design Problem}\label{sec:model}
There is a principal (she) and a finite set of agents, $\{1,\ldots,n\}$, with $n\geq1$. Agent $i$ privately observes his type $t_i$ in a finite nonempty set $T_i$. The type profile $t=(t_1,\ldots,t_n)$ is drawn from a full-support common prior $p\in\Delta(T)^\circ$,\footnote{$\Delta(Y)$ denotes the set of probability distributions on a finite nonempty set $Y$, with $\Delta(Y)^\circ$ denoting those with full support.} where $T\coloneqq\prod_iT_i$, and $p(\cdot\mid t_i)$ denotes the conditional distribution of the other agents' types given $t_i$. The agents' types are \textit{independent} if \(p(t)=\prod_i p_i(t_i)\) for every \(t\in T\), where \(p_i\) is the marginal of \(p\) on \(T_i\).

There is a finite nonempty set of outcomes $X$. 
Agent $i$ receives utility $u_i(x,t)\in\R$ and the principal receives payoff $v(x,t)\in\R$ from outcome $x$ and type profile $t$. All parties evaluate lotteries over outcomes by expected utility: for $\ell\in\Delta(X)$, write $u_i(\ell,t)\coloneqq\sum_x\ell(x)u_i(x,t)$ and $v(\ell,t)\coloneqq\sum_x\ell(x)v(x,t)$. In my main setting, the sets of types $(T_i)_{i=1}^n$, set of outcomes $X$, prior $p$, and payoff functions $(u_i)_{i=1}^n$ and $v$ are common knowledge. An environment has \textit{private values} if each agent's utility depends on the true type profile only through his own type, in which case I write an agent's utility as $u_i(x,t_i)$.

\subsection{Main setting}\label{subsec:mechanisms}

Before the types are drawn, the principal commits to a mechanism. A \textit{finite mechanism} $M$ specifies a finite nonempty set of messages $A_i$ for each agent and an outcome rule $g\colon A\to\Delta(X)$, where $A\coloneqq\prod_iA_i$. Each agent observes the mechanism and his own type before simultaneously sending a message. The outcome is then drawn according to $g(a)$, where $a$ is the profile of messages.

A strategy $\sigma_i\colon T_i\to\Delta(A_i)$ specifies agent $i$'s distribution over messages at each type. Write $\sigma=(\sigma_1,\ldots,\sigma_n)$ for a profile of strategies. I use $t_{-i}$, $a_{-i}$, and $\sigma_{-i}$ for the other agents' types, messages, and strategies. For an agent of type $t_i$ who sends message $a_i$, write his expected utility when the other agents follow $\sigma_{-i}$ as
\[\begin{split} U_i(a_i\mid t_i;M,\sigma_{-i})\coloneqq \sum_{t_{-i}}p(t_{-i}\mid t_i) \sum_{a_{-i}}\left(\prod_{j\ne i}\sigma_j(a_j\mid t_j)\right) u_i\left(g(a_i,a_{-i}),t\right). \end{split}\]

\begin{definition}\label{def:equilibrium}
A \textit{Bayesian Nash equilibrium (BNE)} of $M$ is a profile $\sigma$ satisfying
\begin{equation}\label{eq:bne} \sum_{a_i\in A_i}\sigma_i(a_i\mid t_i)U_i(a_i\mid t_i;M,\sigma_{-i}) \geq U_i(a_i'\mid t_i;M,\sigma_{-i}) \end{equation}
for every agent $i$, every type $t_i\in T_i$, and every message $a_i'\in A_i$. I write $\Eq(M)$ for the set of all BNE.
\end{definition}

An equilibrium is \textit{pure} if every type of every agent chooses one message with probability one. A pure equilibrium is \textit{strict} if each type strictly prefers its prescribed message to every different message, holding fixed the other agents' strategies.

At a profile $\sigma$, the principal's expected payoff before types are drawn is
\begin{equation}\label{eq:principal-expectation} \Pi(M,\sigma)\coloneqq \sum_{t\in T}p(t)\sum_{a\in A} \left(\prod_i\sigma_i(a_i\mid t_i)\right)v(g(a),t). \end{equation}
The principal faces the entire \textit{set of equilibrium payoffs} \(S(M)\coloneqq\left\{\Pi(M,\sigma)\colon\sigma\in\Eq(M)\right\}\), which is nonempty and compact. 

A mechanism is \textit{direct} if $A_i=T_i$ for every agent. I identify a direct mechanism $D$ with its outcome rule $D\colon T\to\Delta(X)$. It is \textit{truthful} if the strategy profile in which each agent always reports his own type is an equilibrium.

Let $\K(\R)$ denote the collection of nonempty compact subsets of $\R$. The principal evaluates a mechanism $M$ according to $\V(S(M))$, where $\V\colon\K(\R)\to\R$.\footnote{\citet{BMV2024} also formulate design objectives as continuous functions of sets of equilibrium outcomes, in an information-design problem. Here the argument of $\V$ is the set of the principal's scalar expected payoffs.} I assume throughout that the evaluation lies between the worst and best possible payoffs:
\begin{equation}\label{eq:internality} \min S\leq\V(S)\leq\max S \qquad\text{for every }S\in\K(\R). \end{equation}
Say that such a \(\V\) is \textit{internal}, and note that this implies $\V(\{s\})=s$. I also assume that $\V$ is continuous in the Hausdorff distance:\footnote{For $x\in\R$ and $S,L\in\K(\R)$, define \(\dist(x,S) \coloneqq\min_{s\in S}|x-s|\) and \(d_H(S,L) \coloneqq \max\left\{\max_{s\in S}\dist(s,L), \max_{\ell\in L}\dist(\ell,S)\right\}\). Two sets are close in this distance when every payoff in either set is close to some payoff in the other.} $d_H(S_m,S)\to0$ implies $\V(S_m)\to\V(S)$.

For an environment $E$, let $\M(E)$ be the set of finite mechanisms, $\D(E)$ its direct mechanisms, and $\D^T(E)$ its truthful direct mechanisms. The principal's unrestricted value and her value when she is restricted to truthful direct mechanisms are
\begin{equation}\label{eq:ordinary-values} V(E)\coloneqq\sup_{M\in\M(E)}\V(S(M)), \qquad \text{\&} \qquad V^T(E)\coloneqq\sup_{D\in\D^T(E)}\V(S(D)). \end{equation}
A constant outcome rule is truthful, so these sets of mechanisms are nonempty. 

\subsection{Unknown prior}\label{subsec:robust-model}

I will also augment the previous setting by assuming the principal does not know the agents' (common) prior as follows. Write \(E^0\coloneqq\left((T_i)_{i=1}^n,X,(u_i)_{i=1}^n,v\right)\) for the type sets, outcomes, and payoff functions, without a prior. Let \(\mathcal{P}\) be a nonempty compact set of full-support distributions on $T\coloneqq\prod_iT_i$. The principal knows $E^0$ and \(\mathcal{P}\) and commits to one mechanism. Nature then selects a prior $p\in\mathcal{P}$, which is common knowledge among the agents. Types are then drawn according to $p$, and each agent observes his own type before the mechanism is played.

The principal first evaluates equilibrium multiplicity conditional on each prior: \(c_p(M)\coloneqq\V(S_p(M))\). Then, for a finite mechanism $M$, define the compact set \(C_{\mathcal{P}}(M) \coloneqq \cl\left\{c_p(M)\colon p\in\mathcal{P}\right\}\). There is then a second functional $\A\colon\K(\R)\to\R$ through which the principal evaluates uncertainty about the prior, so that her overall evaluation of $M$ is \(\A(C_{\mathcal{P}}(M))\). Quite simply, $\V$ describes how the principal treats equilibrium multiplicity conditional on a prior, while $\A$ describes how she treats the set of values generated by the possible priors.

I assume that $\A$ is continuous in the Hausdorff distance and is internal:
\begin{equation}\label{eq:outer-internality} \min K\leq\A(K)\leq\max K \qquad\text{for every }K\in\K(\R). \end{equation}
Like \(\V\), $\A(\{k\})=k$.

I next state a condition on the agents' beliefs after they learn their own types.
\begin{definition}\label{def:common-interim-beliefs}
The set \(\mathcal{P}\) has \textit{common interim beliefs} if \(p(t_{-i}\mid t_i)=q(t_{-i}\mid t_i)\) for every $p,q\in\mathcal{P}$, every agent $i$ with at least two types, every $t_i\in T_i$, and every $t_{-i}\in T_{-i}$.
\end{definition}

A direct mechanism is \textit{robustly truthful} for $(E^0,\mathcal{P})$ if truth telling is an equilibrium under every $p\in\mathcal{P}$. Let $\D^R(E^0,\mathcal{P})$ denote this class. The unrestricted and robustly truthful values are
\[\begin{split} W(E^0,\mathcal{P}) & \coloneqq \sup_{M\in\M(E^0)}\A(C_{\mathcal{P}}(M)),\qquad W^R(E^0,\mathcal{P})\coloneqq \sup_{D\in\D^R(E^0,\mathcal{P})} \A(C_{\mathcal{P}}(D)). \end{split}\]

\section{The Characterizations}\label{sec:main}

First, a restriction to truthful direct mechanisms preserves the principal's optimal value in every environment precisely when she evaluates every mechanism at its best equilibrium:
\begin{theorem}\label{thm:ordinary}
The following conditions are equivalent:
\begin{enumerate}[nosep,noitemsep]
\item\label{cond:ordinary-i} $V^T(E)=V(E)$ for every finite environment $E$.
\item\label{cond:ordinary-ii} In every finite environment $E$, for every $M\in\M(E)$ there is a $D\in\D^T(E)$ such that $\V(S(D))\geq\V(S(M))$.
\item\label{cond:ordinary-iii} $\V(S)=\max S$ for every $S\in\K(\R)$.
\end{enumerate}
If these conditions fail, there is an environment with two agents whose types are independent and uniformly distributed, and a finite mechanism $M^+$ with a unique equilibrium, such that
\begin{equation}\label{eq:ordinary-strict-gap} \sup_{D\in\D(E)}\V(S(D)) <\V(S(M^+))=V(E). \end{equation}
The equivalence also holds when conditions~\conditionref{cond:ordinary-i} and~\conditionref{cond:ordinary-ii} are required only for environments with private values and independent types. If the conditions fail, \eqref{eq:ordinary-strict-gap} can also be obtained with two agents, private values, and independent types.
\end{theorem}

Jettisoning the assumption that the principal knows the prior yields my second theorem:

\begin{theorem}\label{thm:robust-revelation}
Suppose that $\A$ is continuous and satisfies \eqref{eq:outer-internality}. The following are equivalent:
\begin{enumerate}[noitemsep,nosep]
\item\label{cond:robust-ordinary-i} $W^R(E^0,\mathcal{P})=W(E^0,\mathcal{P})$ for every finite environment $E^0$ and every nonempty compact set \(\mathcal{P}\) of full-support priors with common interim beliefs.
\item\label{cond:robust-ordinary-ii} For every such $(E^0,\mathcal{P})$ and every $M\in\M(E^0)$, there is a $D\in\D^R(E^0,\mathcal{P})$ such that \(\A(C_{\mathcal{P}}(D))\geq\A(C_{\mathcal{P}}(M))\).
\item\label{cond:robust-ordinary-iii} $\V(S)=\max S$ and $\A(K)=\max K$ for every $S,K\in\K(\R)$.
\end{enumerate}
Suppose $\V(S)=\max S$ and $\A(K)=\max K$. For every nonempty compact set \(\mathcal{P}\) of full-support priors that does not have common interim beliefs, there is a finite environment $E^0$ with the same type sets and private values such that
\begin{equation}\label{eq:robust-sharpness-gap} W^R(E^0,\mathcal{P}) < W(E^0,\mathcal{P})=1. \end{equation}
\end{theorem}
With a known prior, directification can reproduce any selected  equilibrium, and the issue is how the principal evaluates the additional equilibria of the direct mechanism. With an unknown prior, the same direct rule must also satisfy the truth-telling inequalities generated by every possible system of interim beliefs. But common interim beliefs makes those inequalities identical across priors. On the other hand, once that condition fails, there is \textbf{no} attitude toward equilibrium multiplicity or prior uncertainty that can restore a universal revelation principle.

\begin{corollary}\label{cor:common-interim-beliefs-singleton}
Suppose that at least two agents have at least two types. A set of full-support priors has common interim beliefs only if it is a singleton.
\end{corollary}
\begin{proof}
Fix $p,q\in\mathcal{P}$. For every agent $i$ with at least two types, common interim beliefs implies \(\frac{p(t)}{q(t)}= \frac{p_i(t_i)}{q_i(t_i)}\) for every \(t\in T\). If agents $i$ and $j$ both have at least two types, the likelihood ratio $p(t)/q(t)$ is thus simultaneously a function only of $t_i$ and a function only of $t_j$. Because every type profile has strictly positive probability, this ratio must be constant on $T$. The two priors both have total mass one, so the constant is one and $p=q$.
\end{proof}

When at least two agents have at least two types, every nonsingleton compact set of full-support priors admits a private-values environment satisfying \eqref{eq:robust-sharpness-gap}, \textit{even when both evaluations select the maximum.}

\subsection{Richer reports}\label{subsec:full-information}

The ambiguous-prior results are of course something of a consequence of what a ``type'' is and what it means to report it. Namely, although the agents observe the realized prior \(p\), a direct mechanism asks agent \(i\) to report only \(t_i\). What if direct mechanisms may now elicit each agent's full interim information, \((p,t_i)\).

Fix a nonempty compact set \(\mathcal P\) of full-support priors. Because every agent observes \(p\) before sending his message, an equilibrium of a game induced by a mechanism is a measurable family \(\sigma=(\sigma^p)_{p\in\mathcal P}\) such that \(\sigma^p\in\Eq_p(M)\) for every \(p\in\mathcal P\). Let \(\Eq_{\mathcal P}(M)\) denote the set of these equilibria, and each equilibrium \(\sigma\in\Eq_{\mathcal P}(M)\) produces the prior-contingent payoff for the principal \(\pi(M,\sigma)(p)\coloneqq\Pi_p(M,\sigma^p)\) (\(p\in\mathcal P\)).

For each \(\mathcal P\), take an increasing function \(\rho_{\mathcal P}\) that assigns a real number to every bounded measurable function \(f\colon\mathcal P\to\R\).\footnote{By increasing I mean that \(f(p)\leq h(p)\text{ for every }p\in\mathcal P \  \implies  \ \rho_{\mathcal P}(f)\leq\rho_{\mathcal P}(h)\)} I also assume it is normalized for singleton sets of priors: \(\rho_{\{p\}}(f)=f(p)\). The set of values for the principal generated by the equilibria of the game is \(S_{\mathcal P}^{\rho}(M) \coloneqq \cl\left\{ \rho_{\mathcal P}\left(\pi(M,\sigma)\right): \sigma\in\Eq_{\mathcal P}(M) \right\}\), and the principal evaluates \(M\) according to \(\V\left(S_{\mathcal P}^{\rho}(M)\right)\).

For this subsection, a \textit{direct mechanism} endows agent \(i\) the space of messages \(\mathcal P\times T_i\) and has a measurable outcome rule. It is \textit{truthful} if reporting \((p,t_i)\) is an equilibrium under every realized prior \(p\).

\begin{proposition}\label{prop:full-information-revelation}
Let \((\rho_{\mathcal P})_{\mathcal P}\) be increasing and normalized for singleton sets of priors. The following are equivalent:
\begin{enumerate}[nosep,noitemsep]
\item\label{cond:full-information-i} For every finite environment \(E^0\), every nonempty compact set \(\mathcal P\) of full-support priors, and every finite mechanism \(M\), there is a truthful direct mechanism \(D\) such that \(\V\left(S_{\mathcal P}^{\rho}(D)\right)\geq\V\left(S_{\mathcal P}^{\rho}(M)\right)\).
\item\label{cond:full-information-ii} \(\V(S)=\max S\) for every \(S\in\K(\R)\).
\end{enumerate}
\end{proposition}

\begin{proof} Necessity of \(\max\) is immediate from  \Cref{thm:ordinary} by taking \(\mathcal P=\{p\}\). Sufficiency is an easy consequence of the classical revelation principle.
\end{proof}

\section{Known prior proof}\label{sec:construction}

For any finite set $F$ containing at least two payoffs for the principal, I construct an environment with a truthful direct mechanism whose set of equilibrium payoffs is exactly $F$. Adding one additional message for each agent eliminates the inferior equilibria. Moreover, any other direct mechanism that attains the best payoff must reproduce the entire set $F$. Strict equilibria then ensure that direct mechanisms approaching that best payoff also have equilibria with payoffs close to every member of $F$.\footnote{\citet{LSV2011} study which sets can arise as equilibrium payoff sets of finite games.}

A \textit{relabeling} of a direct mechanism $D$ permutes the labels of each agent's reports. Formally, for bijections $\pi_i\colon T_i\to T_i$, the relabeled mechanism assigns the lottery $D(\pi_1^{-1}(r_1),\ldots,\pi_n^{-1}(r_n))$ to the report profile $(r_1,\ldots,r_n)$.
\begin{lemma}\label{lem:construction}
Let $F=\{a_0,a_1,\ldots,a_k\}$, with $b\coloneqq a_0>a_1>\cdots>a_k$ and $k\geq1$. There is an environment with two agents and independent uniform types, a truthful direct mechanism $D^0$, and a mechanism $M^+$ with one additional message for each agent, satisfying:
\begin{enumerate}[noitemsep,nosep]
\item At every type profile, the principal's payoff is at most $b$, and exactly one outcome gives her $b$. These maximizing outcomes differ across type profiles.
\item $S(D^0)=F$, and every member of $F$ is attained at a strict pure equilibrium of $D^0$.
\item $M^+$ has a unique BNE, with $S(M^+)=\{b\}$.
\item Every direct mechanism $D$ with $b\in S(D)$ is a relabeling of $D^0$. 
\end{enumerate}
\end{lemma}

In this lemma, proved in \Cref{app:construction}, I flesh out the construction in \secref{sec:nec}. An implication of it is that any direct mechanism $D$ with $b\in S(D)$ is a relabeling of $D^0$ and so $S(D)=F$. The next lemma (proved in \Cref{app:nearmax}) also controls direct mechanisms whose best equilibrium payoffs are merely close to $b$. This will be needed because the values in \eqref{eq:ordinary-values} are suprema.

\begin{lemma}\label{lem:near-maximum}
In the environment of \Cref{lem:construction}, for every $\eta>0$ there is a $\delta>0$ such that every direct mechanism $D$ satisfies
\begin{equation}\label{eq:near-maximum} \max S(D)>b-\delta \quad \implies \quad d_H(S(D),F)<\eta. \end{equation}
\end{lemma}

\begin{lemma}\label{lem:revelation}
For every finite mechanism $M$ and every $\sigma\in\Eq(M)$, there is a truthful direct mechanism whose truthful equilibrium induces the same outcome distribution as $(M,\sigma)$ at every type profile.
\end{lemma}
\begin{proof}
This is the standard revelation principle construction a la \citet{Myerson1982}.
\end{proof}

\subsection{Proof of \texorpdfstring{\Cref{thm:ordinary}}{the theorem}}\label{subsec:ordinary-proof}

\begin{proof}[Proof of \Cref{thm:ordinary}]
Suppose first that $\V(S)=\max S$ for every compact set. Fix $M$ and choose an equilibrium attaining $\max S(M)$. By \Cref{lem:revelation}, there is a truthful direct mechanism $D$ whose truthful equilibrium gives the principal that same payoff. Consequently, \(\V(S(D))=\max S(D)\geq\max S(M)=\V(S(M))\), proving that \conditionref{cond:ordinary-iii} implies \conditionref{cond:ordinary-ii}. \conditionref{cond:ordinary-ii} implies \conditionref{cond:ordinary-i} by taking suprema and using $\D^T(E)\subseteq\M(E)$.

For necessity, suppose $\V(S)<\max S$ for some $S\in\K(\R)$, and write $b=\max S$. Choose finite subsets $F_m\subseteq S$ containing $b$ such that $d_H(F_m,S)\to0$.\footnote{Such sets exist because $S$ is compact: for each $m$, take finitely many points of $S$ within distance $1/m$ of every point of $S$, and add $b$.} Continuity of \(\V\) implies $\V(F_m)\to\V(S)<b$. Fix one such finite set $F$ with $\V(F)<b$. Since $\V(\{b\})=b$, this set has at least two members.

Use the environment and mechanisms from \Cref{lem:construction} for $F$. Set $\gamma\coloneqq(b-\V(F))/2>0$. By continuity of \(\V\) at $F$, there is an $\eta>0$ such that
\begin{equation}\label{eq:continuity-at-finite-set} d_H(L,F)<\eta\quad \implies \quad\V(L)<b-\gamma. \end{equation}
Let $\delta>0$ be given by \Cref{lem:near-maximum} for this $\eta$. Every direct mechanism $D$ falls into one of two cases. If $\max S(D)\leq b-\delta$, then \eqref{eq:internality} implies $\V(S(D))\leq b-\delta$. If $\max S(D)>b-\delta$, then \eqref{eq:near-maximum} and \eqref{eq:continuity-at-finite-set} imply $\V(S(D))<b-\gamma$. Hence,
\begin{equation}\label{eq:uniform-direct-gap} \sup_{D\in\D(E)}\V(S(D)) \leq b-\min\{\delta,\gamma\}<b. \end{equation}
The principal's payoff $v(x,t)$ never exceeds $b$, while $M^+$ has the singleton payoff set $\{b\}$. Thus, $V(E)=\V(S(M^+))=b$. \eqref{eq:uniform-direct-gap} proves \eqref{eq:ordinary-strict-gap} and shows that \conditionref{cond:ordinary-i} fails. 
I prove the assertions concerning private values in \secref{sec:private-values}.\end{proof}

\section{Unknown prior proof}\label{subsec:robust-proof}

\begin{lemma}\label{lem:prior-upper-semicontinuity}
Fix a finite environment $E^0$. For every finite mechanism $M$, the function \(p\mapsto \max S_p(M)\) is upper semicontinuous on the set of full-support priors.
\end{lemma}

I prove this lemma in \Cref{appuppersemi}.

\begin{lemma}\label{lem:robust-directification}
Suppose that \(\mathcal{P}\) has common interim beliefs. Fix $p^*\in\mathcal{P}$, a mechanism $M$ with outcome rule $g$, and an equilibrium $\sigma^*\in\Eq_{p^*}(M)$. The direct mechanism \(D(\left.x\right|r) \coloneqq \sum_{a\in A}\left(\prod_i \sigma^*_i(a_i|r_i)\right)g(a)(x)\) (\(x \in X\), \(r \in T\)) is truthful under every $p\in\mathcal{P}$.
\end{lemma}
I prove this in \Cref{robustdirectproof}. The next two lemmata identify the principal's evaluation of uncertainty over priors and the necessity of common interim beliefs, again filling in the gaps present in \secref{sec:nec}. See \Cref{app:outer,app:sharp}.

\begin{lemma}\label{lem:outer-construction}
Let \(F=\{b,a_1,\ldots,a_k\}\), where \(k \geq 1\), \(b=\max F\) and \(a_j<b\) for every \(j\). There is a finite environment \(E^0\) and a finite set \(\mathcal P\) of full-support priors with common interim beliefs such that the following statements hold for every internal \(\V\) and every continuous internal \(\A\).
\begin{enumerate}[nosep,noitemsep]
\item The optimal unrestricted value equals \(b\).
\item Every robustly truthful direct mechanism generates, for some \(q\in[0,1]\), the set of prior-specific values
\begin{equation}\label{eq:outer-direct-frontier} K_q\coloneqq \left\{ qb+(1-q)a_j,\, qa_j+(1-q)b\colon j=1,\ldots,k \right\}. \end{equation}
Conversely, every \(q\in[0,1]\) is attained by a robustly truthful direct mechanism.
\end{enumerate}
\end{lemma}

\begin{lemma}\label{lem:common-interim-beliefs-sharp}
Fix a nonempty compact set \(\mathcal{P}\) of full-support priors that does not have common interim beliefs. There is a finite environment $E^0$ with the same type sets and private values and a number $\bar v<1$ such that, when $\V(S)=\max S$ and $\A(K)=\max K$, \(W(E^0,\mathcal{P})=1\), while \(W^R(E^0,\mathcal{P}) \leq\bar v\).
\end{lemma}

\begin{proof}[Proof of \Cref{thm:robust-revelation}]
I begin with sufficiency. Suppose that $\V(S)=\max S$ and $\A(K)=\max K$ for every compact set. Fix $(E^0,\mathcal{P})$ satisfying common interim beliefs and fix $M\in\M(E^0)$. The principal assigns $M$ the value \(\A(C_{\mathcal{P}}(M)) =\max_{p\in\mathcal{P}}\max S_p(M)\). By \Cref{lem:prior-upper-semicontinuity} and compactness of \(\mathcal{P}\), some $p^*\in\mathcal{P}$ attains this maximum. Choose an equilibrium $\sigma^*\in\Eq_{p^*}(M)$ attaining $\max S_{p^*}(M)$. By \Cref{lem:robust-directification}, the direct mechanism $D$ obtained from $\sigma^*$ is truthful under every prior in \(\mathcal{P}\). Its truthful equilibrium under $p^*$ delivers the principal $\max S_{p^*}(M)$. Because both $\V$ and $\A$ select the maximum, $\A(C_{\mathcal{P}}(D))\geq\A(C_{\mathcal{P}}(M))$.

I next prove necessity. A singleton set of priors satisfies common interim beliefs, and \eqref{eq:outer-internality} implies $\A(\{c\})=c$. Accordingly, the equivalence in \Cref{thm:ordinary} implies that universal robust revelation requires $\V(S)=\max S$ for every compact set.

It remains to identify \(\A\). Fix a finite set \(F=\{b,a_1,\ldots,a_k\}\), where \(b=\max F\), \(a_j<b\), and \(k\geq1\), and apply \Cref{lem:outer-construction}. The unrestricted value is \(b\). Suppose \(\A(F)<b\). When \(q=0\) or \(q=1\), the set in \eqref{eq:outer-direct-frontier} equals \(F\), so the principal assigns it a value below \(b\). For \(q\in(0,1)\), all prior-specific values are strictly below \(b\), so \eqref{eq:outer-internality} implies that she again assigns the set a value below \(b\). Because \(\A\) is continuous, her evaluation varies continuously with \(q\) and attains a maximum on \([0,1]\). This maximum is strictly below \(b\), since her evaluation is below \(b\) at every \(q\). Consequently, equality of the unrestricted and robustly truthful values requires \(\A(F)=\max F\) for every finite set \(F\) with at least two members. The same conclusion holds for singletons by \eqref{eq:outer-internality}. For every \(K\in\K(\R)\), choose finite sets \(F_m\subseteq K\) with \(d_H(F_m,K)\to0\). Continuity of \(\A\) and of the maximum then implies \(\A(K)=\lim_m\A(F_m)=\lim_m\max F_m=\max K\). This proves the equivalence.

Finally, \Cref{lem:common-interim-beliefs-sharp} delivers \eqref{eq:robust-sharpness-gap} for every compact full-support prior set that violates common interim beliefs.
\end{proof}

\section{Unrestricted transfers}\label{sec:unrestricted-transfers}

I now allow the principal to choose monetary transfers freely. As we will see, the revelation principle no longer requires optimism. Retain the finite type sets $T_i$ and private values, and suppose that the agents' types are independent under the known full-support prior $p$. At outcome $x$, type profile $t$, and payments $\tau=(\tau_1,\ldots,\tau_n)$ made by the agents, agent $i$ receives utility $u_i(x,t_i)-\tau_i$, and the principal receives $v(x,t)+\sum_i\tau_i$. I assume that, for every agent $i$ and distinct types $t_i,s_i\in T_i$, the function $x\mapsto u_i(x,t_i)-u_i(x,s_i)$ is not constant on $X$.

For this section, a finite mechanism $M=((A_i)_i,q,\tau)$ specifies finite nonempty message sets, an allocation rule $q\colon A\to\Delta(X)$, and payments $\tau_i\colon A\to\R$. A direct mechanism has $A_i=T_i$ for every agent. Write $\sigma^T$ for truthful reporting, and call a direct mechanism \textit{strictly Bayesian incentive compatible (strictly BIC)} if every type strictly prefers truth to every other report when the other agents report truthfully.

For a strategy profile $\sigma$, let $q^\sigma(x\mid t)$ be the probability of outcome $x$ conditional on the true type profile $t$, and let $\bar\tau_i^\sigma(t_i)$ be type $t_i$'s interim expected payment. Write $U_i(t_i;M,\sigma)$ for his interim expected net utility. The principal's expected profit is
\[\Pi(M,\sigma)=\sum_{t\in T}p(t)\sum_{x\in X}q^\sigma(x\mid t)v(x,t)+\sum_i\sum_{t_i\in T_i}p_i(t_i)\bar\tau_i^\sigma(t_i).\]
As before, $\Eq(M)$ contains all BNEs and $S(M)=\{\Pi(M,\sigma)\colon\sigma\in\Eq(M)\}$. 

Agents have type-dependent outside options: $o_i(t_i)\in\R$ is type $t_i$'s outside option (in utils). An equilibrium is \textit{interim individually rational (interim IR)} if
\begin{equation}\label{eq:ut-ir} U_i(t_i;M,\sigma)\geq o_i(t_i) \qquad\text{for every }i\text{ and }t_i\in T_i. \end{equation}
Let $\M(E)$ denote the finite mechanisms in this quasilinear environment, let $\M^I(E)$ contain those mechanisms whose every equilibrium satisfies \eqref{eq:ut-ir}, and let $\D^T(E)$ be its truthful direct members. Write
\[\Pi^{*}(E)\coloneqq\sup\left\{\Pi(M,\sigma)\colon M\in\M(E),\ \sigma\in\Eq(M)\text{ satisfies \eqref{eq:ut-ir}}\right\}.\]
\begin{proposition}\label{prop:ut-revelation}
For every finite mechanism $M$, every interim-IR equilibrium $\sigma\in\Eq(M)$, and every $\varepsilon>0$, there is a truthful direct mechanism $D_\varepsilon$ whose unique BNE is interim IR and satisfies \(\left|\Pi(D_\varepsilon,\sigma^T)-\Pi(M,\sigma)\right|<\varepsilon\).
\end{proposition}

The upshot is that for every internal $\V$, \[\sup_{D\in\D^T(E)}\V(S(D))=\sup_{M\in\M^I(E)}\V(S(M))=\Pi^{*}(E).\]
Others\footnote{E.g., \citet{AbreuMatsushima1992Incomplete}, \citet{dAspremontCremerGerardVaret2005}, and \citet{Kim2013}.} obtain their implementation results by enlarging agents' message spaces, while \citet{OllarPenta2017} obtain full implementation in direct mechanisms with a particular smooth structure.\footnote{Differentiable valuation, allocation, and transfer functions, strictly concave expected payoffs in each agent's own report, and bounds on how opponents' reports affect incentives.} I generate a unique truthful BNE from any strictly BIC direct mechanism with finite type spaces, independent types, and private values.

\begin{lemma}\label{lem:ut-transfer-modification}
For any strictly BIC direct mechanism $(q,\tau)$, there is a payment rule $\hat\tau$ such that $(q,\hat\tau)$ has truth telling as its unique BNE and
\begin{equation}\label{eq:ut-preserved-payments} \sum_{t_{-i}}p_{-i}(t_{-i})\hat\tau_i(r_i,t_{-i})=\sum_{t_{-i}}p_{-i}(t_{-i})\tau_i(r_i,t_{-i}) \qquad\text{for every }i\text{ and }r_i\in T_i. \end{equation}
\end{lemma}

I prove this lemma and \Cref{prop:ut-revelation} in \Cref{app:propproof}.

\appendix
\section{Omitted proofs}

\subsection{Proof of \texorpdfstring{\Cref{lem:construction}}{lemma}}\label[appendix]{app:construction}
\begin{proof}[Proof of \Cref{lem:construction}]
Consider the following environment and mechanism. Let $N\coloneqq k+1$ and $T_1=T_2=\{0,\ldots,k\}$. The agents' types are independent, and each is uniformly distributed on $\{0,\ldots,k\}$. Let $C$ be a symbol outside the sets of types, and let $X\coloneqq\left(T_1\cup\{C\}\right) \times\left(T_2\cup\{C\}\right)$. Write $x=(x_1,x_2)$ for an outcome. The direct mechanism $D^0$ chooses outcome $x=(r_1,r_2)$ with probability one when the agents report $(r_1,r_2)$.

For $x_i\in T_i$, let $j_i$ be the unique integer in $\{0,\ldots,k\}$ congruent to $x_i-t_i$ modulo $N$. If $x_i=C$, let $j_i=C$, so that $j_i=0$ if and only if $x_i=t_i$. Under $D^0$, a type $t_i$ can choose any difference $j_i=j$ by reporting $(t_i+j)\bmod N$.

For $x_1\in T_1$ and $x_2\in T_2$, give agent $i$ utility \(u_i(x,t)=-\left(j_i-j_{-i}\right)^2\). If $x_1=C$ or $x_2=C$, define agent $i$'s utility by $u_i(x,t)=\tilde{u}_i(j_i,j_{-i})$, where
\[\tilde{u}_i(C,0) = -\frac12 = \tilde{u}_i(C,C), \quad \tilde{u}_i(0,C) = 0, \quad \text{\&} \quad -\tilde{u}_i(j,C) = \tilde{u}_i(C,j) = 1 \quad (j=1,\ldots,k).\]

I choose the principal's payoffs so that equilibria with a common difference attain the values in $F$, while the mixed equilibria don't introduce additional payoffs. To that end, define a symmetric matrix $W$ by
\begin{equation}\label{eq:principal-matrix} W_{jj} = a_j \quad (j=0,\ldots,k),\quad \text{\&} \quad W_{j,j+1}=W_{j+1,j} =\frac{3a_{j+1}-a_j}{2} \quad (j=0,\ldots,k-1),\end{equation}
and set every other off-diagonal entry to $a_k$. If $x_i\in T_i$ for both agents, let $v(x,t)=W_{j_1j_2}$; otherwise, let $v(x,t)=a_k$.

Only $W_{00}$ equals $b$. Every other diagonal entry is below $b$, and the adjacent entries satisfy $(3a_{j+1}-a_j)/2<a_{j+1}<b$. Thus, the principal's payoff never exceeds $b$, and the unique outcome attaining $b$ at type profile $t$ is $x=t$. Distinct type profiles have distinct maximizing outcomes. This establishes part~\textnormal{(i)}.

\begin{claim}\label{claim:direct-equilibria}
The direct mechanism has $S(D^0)=F$, and every member of $F$ is attained at a strict pure equilibrium.
\end{claim}
\begin{proof}
Fix a strategy profile under $D^0$. Let $q_i(j)$ be the unconditional probability that agent $i$'s difference equals $j$:
\[q_i(j)\coloneqq\frac1N\sum_{t_i=0}^k \sigma_i\left((t_i+j)\bmod N\mid t_i\right) \qquad (j=0,\ldots,k).\]
Independence of types implies that every type of agent $i$ faces the same distribution $q_{-i}$ of the other agent's difference. Because each type can choose every difference, in equilibrium every difference assigned strictly positive probability by $q_i$ must maximize expected utility against $q_{-i}$. Conversely, any equilibrium of the finite game in which the two agents choose differences and receive utility $-(j_i-j_{-i})^2$ can be induced in $D^0$: each type uses that equilibrium distribution over differences.

Against an opponent whose difference has mean $\mu$, the expected utility from choosing difference $j$ is \(-j^2+2j\mu-\E[j_{-i}^2]\). The best replies are the integers closest to $\mu$. If one agent chooses a pure difference $j$, the other's unique best reply is $j$, so both choose that difference. If both distributions are nondegenerate, each must assign strictly positive probability on exactly two adjacent integers. Suppose the first agent uses $\{j,j+1\}$ and the second uses $\{\ell,\ell+1\}$. The first agent's indifference requires the second agent's mean to be $j+1/2$, which mean lies strictly between $\ell$ and $\ell+1$, so $j=\ell$. The second agent's indifference then requires the first agent's mean also to equal $j+1/2$. Hence, both agents assign probability $1/2$ to each of $j$ and $j+1$.

This enumerates all possible pairs $(q_1,q_2)$ induced by  equilibria (including equilibria in which different types use different mixtures). At a common pure difference $j$, the principal receives $a_j$. At common equal mixtures over $j$ and $j+1$, the two differences are independent because the types and private randomizations are independent. Thus, the principal's expected payoff is
\begin{equation}\label{eq:mixed-principal-payoff} \frac14a_j+\frac14a_{j+1} +\frac12\frac{3a_{j+1}-a_j}{2}=a_{j+1}. \end{equation}
Every equilibrium payoff consequently belongs to $F$.

For each $j\in\{0,\ldots,k\}$, have every type of both agents report $(t_i+j)\bmod N$. Each type obtains utility $0$, whereas every different report yields utility at most $-1$. This is a strict pure equilibrium, and the principal receives $a_j$. These equilibria attain every member of $F$, establishing the claim. The case $j=0$ is truth telling, so $D^0$ is truthful.
\end{proof}

\begin{claim}\label{claim:unique-indirect} The mechanism $M^+$, which provides each agent $i$ with the message set $T_i\cup\{C\}$ and chooses $x=a$ with probability one after the message profile $a$, has a unique equilibrium, in which both agents report truthfully.\end{claim}
\begin{proof}
The game is dominance solvable: first, for every agent and type, \(C\) strictly dominates a false report. Second, after eliminating the false reports, truthful reporting strictly dominates \(C\).
\end{proof}

\begin{claim}\label{claim:all-direct-rules}
Every direct mechanism with an equilibrium giving the principal $b$ is a relabeling of $D^0$.
\end{claim}
\begin{proof}
This follows immediately from the fact that if the principal obtains \(b\), for every type profile $t$, every report profile used with strictly positive probability must lead to the unique maximizing outcome $x=t$ with probability one.\end{proof}

\Cref{claim:direct-equilibria,claim:unique-indirect,claim:all-direct-rules} establish parts~\textnormal{(ii)}, \textnormal{(iii)}, and~\textnormal{(iv)}, respectively.
\end{proof}

\subsection{Proof of \texorpdfstring{\Cref{lem:near-maximum}}{lemma}}\label[appendix]{app:nearmax}
\begin{proof}[Proof of \Cref{lem:near-maximum}]
Suppose the conclusion fails for some $\eta>0$. Then there are direct mechanisms $D_m$ satisfying \(\max S(D_m)>b-\frac1m\) and \(d_H(S(D_m),F)\geq\eta\). For each $m$, choose an equilibrium $\sigma_m$ attaining $\max S(D_m)$. The space of direct rules is the finite product $\Delta(X)^T$, and the space of strategy profiles is also a finite product of simplices. Both are compact. After passing to a subsequence, write $D_m\to D^*$ and $\sigma_m\to\sigma^*$.

The equilibrium inequalities \eqref{eq:bne} are continuous in the outcome rule and strategies, thus hold at the limit, so $\sigma^*\in\Eq(D^*)$. Since $(D,\sigma)\mapsto\Pi(D,\sigma)$ is continuous and \(b-\frac1m<\Pi(D_m,\sigma_m)\leq b\), \(\Pi(D^*,\sigma^*)=b\). Part~\textnormal{(iv)} of \Cref{lem:construction} implies that $D^*$ is a relabeling of $D^0$, so $S(D^*)=F$. Every payoff in $F$ is attained at a strict pure equilibrium of $D^*$.

\begin{claim}\label{claim:hausdorff-limit}
$d_H(S(D_m),F)\to0$.
\end{claim}
\begin{proof}
Fix $a\in F$ and a strict pure equilibrium of $D^*$ attaining $a$. The finitely many strict inequalities saying that each type prefers its prescribed message continue to hold when the outcome rule is sufficiently close to $D^*$. Thus, the same pure strategy profile is an equilibrium of $D_m$ for all sufficiently large $m$. Its principal payoff converges to $a$. Since $F$ is finite, this proves \(\max_{a\in F}\dist(a,S(D_m))\rightarrow 0\).

For the other direction, suppose there are equilibrium payoffs of $D_m$ that remain at least some $\varepsilon>0$ from $F$ along a subsequence. Choose $\tau_m\in\Eq(D_m)$ attaining such a payoff. After passing to a further subsequence, write $\tau_m\to\tau^*$. Passing to the limit in \eqref{eq:bne} yields $\tau^*\in\Eq(D^*)$. Since $(D,\tau)\mapsto\Pi(D,\tau)$ is continuous, \(\Pi(D_m,\tau_m)\to\Pi(D^*,\tau^*)\in F\), a contradiction. Therefore, \(\max_{s\in S(D_m)}\dist(s,F)\rightarrow 0\). \end{proof}

\Cref{claim:hausdorff-limit} contradicts $d_H(S(D_m),F)\geq\eta$, completing the proof.
\end{proof}

\subsection{Proof of \texorpdfstring{\Cref{lem:prior-upper-semicontinuity}}{}}\label[appendix]{appuppersemi}
\begin{proof}[Proof of \Cref{lem:prior-upper-semicontinuity}]
Fix a mechanism $M$ and a sequence of full-support priors $p_m\to p$. For each $m$, choose $\sigma_m\in\Eq_{p_m}(M)$ attaining $\max S_{p_m}(M)$. Along a subsequence realizing the upper limit, compactness of the strategy space implies $\sigma_m\to\sigma$. The equilibrium inequalities depend continuously on the prior and the strategy profile on the full-support simplex, so $\sigma\in\Eq_p(M)$. The principal's expected payoff is also continuous. Therefore, \(\limsup_m\max S_{p_m}(M) =\Pi_p(M,\sigma) \leq\max S_p(M)\).\end{proof}

\subsection{Proof of \texorpdfstring{\Cref{lem:robust-directification}}{}}\label[appendix]{robustdirectproof}
\begin{proof}[Proof of \Cref{lem:robust-directification}]
Fix an agent $i$ with at least two types, a type $t_i$, and a possible report $r_i$. Holding the other agents to truthful reports in the direct mechanism, the difference between the expected utilities from reporting $t_i$ and $r_i$ can be written as \(\sum_{t_{-i}}p(t_{-i}\mid t_i) \Delta_i(t_{-i};t_i,r_i)\), where $\Delta_i$ is determined by the outcome rule of $M$ and the fixed message distributions in $\sigma^*$. Under $p^*$, this expression is nonnegative because $\sigma^*$ is an equilibrium of $M$. Common interim beliefs makes the conditional probabilities in the display identical under every $p\in\mathcal{P}$. The same truth-telling inequality therefore holds under every possible prior. An agent with one type has no alternative direct report. Thus, the direct mechanism is robustly truthful.
\end{proof}

\subsection{Proof of \texorpdfstring{\Cref{lem:outer-construction}}{lemma}}\label[appendix]{app:outer}
\begin{proof}[Proof of \Cref{lem:outer-construction}]
There are two agents. Agent~1 has the singleton type set \(T_1=\{*\}\), and agent~2 has type set \(T_2=\{0,1,2\}\). Choose \(L<\min F\) and \(H>2b-L\). For each \(j=1,\ldots,k\), let \(\beta\coloneqq(b-L)/(H-L)\) and \(\alpha_j\coloneqq(a_j-L)/(H-L)\), and define \(p_j^0\coloneqq(\beta,\alpha_j,1-\beta-\alpha_j)\) and \(p_j^1\coloneqq(\alpha_j,\beta,1-\beta-\alpha_j)\). Let \(\mathcal{P}\coloneqq\{p_j^0,p_j^1\colon j=1,\ldots,k\}\). The only agent with more than one type has the same degenerate belief about agent~1 under every prior, so \(\mathcal{P}\) has common interim beliefs.

There are two outcomes, \(X=\{0,1\}\). At types \(0\), \(1\), and \(2\) of agent~2, respectively, agent~1 and the principal receive the outcome-payoff pairs \((H,L)\), \((L,H)\), and \((L,L)\). Agent~2 receives utility one from outcome zero and zero from outcome one.

Consider the indirect mechanism \(M^*\). Agent~1 chooses a message in \(\{0,1\}\), agent~2 has one message, and the outcome equals agent~1's message. Under \(p_j^0\), outcomes zero and one give agent~1 and the principal the expected payoffs \(b\) and \(a_j\), respectively. Under \(p_j^1\), they give the expected payoffs \(a_j\) and \(b\). The mechanism has a unique equilibrium under every prior and delivers the principal \(b\). Hence, \(C_{\mathcal{P}}(M^*)=\{b\}\) and \(\A(C_{\mathcal{P}}(M^*))=b\).

No mechanism can produce for the principal an equilibrium payoff above \(b\) under any prior in \(\mathcal{P}\), as every type of agent~2 has the same preferences over outcomes and so there must be a common probability \(q\) of outcome zero following agent~2's report. Consequently, for a robustly truthful direct mechanism, the principal's equilibrium payoffs under \(p_j^0\) and \(p_j^1\) are \(qb+(1-q)a_j\) and \(qa_j+(1-q)b\). Their equilibrium-payoff sets are singletons, so \eqref{eq:outer-direct-frontier} is also the set of prior-specific values. Every \(q\in[0,1]\) is attained by choosing outcome zero with probability \(q\) after every report.
\end{proof}

\subsection{Proof of \texorpdfstring{\Cref{lem:common-interim-beliefs-sharp}}{lemma}}\label[appendix]{app:sharp}
\begin{proof}[Proof of \Cref{lem:common-interim-beliefs-sharp}]
Because common interim beliefs fails, there are $p,q\in\mathcal{P}$, an agent $i$ with at least two types, and a type $t_i^0$ such that \(p(\cdot\mid t_i^0)\ne q(\cdot\mid t_i^0)\). Choose a function $d\colon T_{-i}\to\R$ satisfying
\begin{equation}\label{eq:separating-interim-beliefs} \sum_{t_{-i}}p(t_{-i}\mid t_i^0)d(t_{-i})>0> \sum_{t_{-i}}q(t_{-i}\mid t_i^0)d(t_{-i}). \end{equation}
Such a function exists because the two conditional distributions are distinct: choose an event to which they assign different probabilities and subtract a constant strictly between those probabilities.

Take $X=T$ and give the principal payoff \(v(x,t)=\mathbf 1_{\{x=t\}}\). For every agent $j\ne i$, set \(u_j(x,t_j)=\mathbf 1_{\{x_j=t_j\}}\), and for agent $i$, set
\[u_i(x,t_i)= \begin{cases} d(x_{-i}),&t_i=t_i^0\text{ and }x_i=t_i^0,\\ 0,&t_i=t_i^0\text{ and }x_i\ne t_i^0,\\ \mathbf 1_{\{x_i=t_i\}},&t_i\ne t_i^0. \end{cases}\]
Let $D^0$ choose outcome $x=r$ for sure when the agents report $r\in T$. Truth telling is a strict equilibrium under prior $p$ as type $t_i^0$ of agent $i$ strictly prefers truth by the first inequality in \eqref{eq:separating-interim-beliefs} and every other type receives utility one from reporting truthfully and zero from any false report. The principal receives one, and no mechanism can exceed the upper bound one, so the unrestricted value is one.

Consider a direct mechanism $D$ with an equilibrium giving the principal one under some full-support prior. At every type profile $t$, $D$ must assign probability one to $x=t$ after every report profile that the agents use with strictly positive probability. By the argument in the proof of \Cref{claim:all-direct-rules}, each type uses one report, and distinct types of an agent use distinct reports. Let each agent $j$ report according to $\pi_j\colon T_j\to T_j$ in this equilibrium. Each $\pi_j$ is a bijection, and after reports $r$, $D$ assigns probability one to the outcome $x=\left(\pi_1^{-1}(r_1),\ldots,\pi_n^{-1}(r_n)\right)$.

Suppose $D$ is truthful under both $p$ and $q$. Every type $t_j$ of every agent $j$, except for type $t_i^0$ of agent $i$, receives utility one from reporting $\pi_j(t_j)$ and zero from any other report, regardless of the other agents' reports. Truthfulness thus requires $\pi_j(t_j)=t_j$ for all these types. Since $\pi_i$ is a bijection and maps every other type to itself, it must also map $t_i^0$ to itself. So, $D=D^0$. Under $q$, however, type $t_i^0$ receives strictly negative expected utility from reporting truthfully when the other agents report truthfully, by the second inequality in \eqref{eq:separating-interim-beliefs}. Any false report gives him zero, so $D$ cannot be truthful under both priors.

Let $\Gamma$ be the set of triples $(D,r,\sigma)$ such that $D$ is a direct rule truthful under both $p$ and $q$, $r\in\mathcal{P}$, and $\sigma\in\Eq_r(D)$. This set is compact.\footnote{The space of direct rules and the strategy space are compact, truthfulness under $p$ and $q$ is given by finitely many closed inequalities, \(\mathcal{P}\) is compact, and the equilibrium inequalities have a closed graph in $(D,r,\sigma)$.} No triple in $\Gamma$ delivers the principal payoff one. The principal's expected payoff $\Pi_r(D,\sigma)$ is continuous in $(D,r,\sigma)$, so it has a maximum $\bar v<1$ on $\Gamma$. Every robustly truthful direct mechanism is truthful under $p$ and $q$, so all of its equilibrium payoffs under every $r\in\mathcal{P}$ are at most $\bar v$, and so $W^R(E^0,\mathcal{P})\leq\bar v$.\end{proof}

\subsection{Proofs of \texorpdfstring{\Cref{prop:ut-revelation} and \Cref{lem:ut-transfer-modification}}{results}}\label[appendix]{app:propproof}

\begin{proof}[Proof of \Cref{lem:ut-transfer-modification}]
If at most one agent has more than one type, strict IC already makes truth telling the unique equilibrium, so take $\hat\tau=\tau$. Otherwise, for each agent $i$, fix an arbitrary report $t_i^0\in T_i$. For each $a\in T_i\setminus\{t_i^0\}$, define \(\phi_{ia}(r_i)\coloneqq\mathbf1_{\{r_i=a\}}-p_i(a)\).

I now modify agent $i$'s payment from $\tau_i(r)$ to $\tau_i(r)-\kappa_iH_i(r)$ as follows. I choose $H_i$ so that, for every report $r_i$, agent \(i\)'s expected payment when his opponents report truthfully is the same under $\tau_i-\kappa_iH_i$ as under $\tau_i$. As the original mechanism was strictly BIC, he strictly prefers to be truthful when his opponents are. Crucially, the modifications may affect his incentives when his opponents do not report truthfully, and I will use this to eliminate the other equilibria.

Now let us construct these objects. Start by initially setting $H_i(r)=0$ for every agent and report profile. For each agent $i$, consider the reports in $T_i\setminus\{t_i^0\}$. Choose two agents $i<j$ and one of these reports from each, say $a$ and $b$. Add $\phi_{ia}(r_i)\phi_{jb}(r_j)$ to $H_i(r)$ and subtract it from $H_j(r)$. Repeat this procedure without choosing the same report of the same agent twice, until no further pair can be chosen.

If any reports remain unused in this construction, they must all belong to one agent, as otherwise, I could choose another pair. Denote that agent by $i_0$. If none remain, choose any agent with more than one type as $i_0$. Each product is added to one function and subtracted from another, so \(\sum_i H_i(r)=0\) for every \(r\).

Every term in $H_i(r)$ is either $\phi_{ia}(r_i)\phi_{jb}(r_j)$ or its negative, for some $j\neq i$. Fix $r_i$ and suppose the opponents report truthfully. Then:
\[
\sum_{t_{-i}}p_{-i}(t_{-i})\phi_{ia}(r_i)\phi_{jb}(t_j)
=\phi_{ia}(r_i)\sum_{t_j}p_j(t_j)
\left[\mathbf 1_{\{t_j=b\}}-p_j(b)\right]
=0.
\]
In turn, fix $r_{-i}$ and suppose agent $i$ reports truthfully, in which case
\[
\sum_{t_i}p_i(t_i)\phi_{ia}(t_i)\phi_{jb}(r_j)
=\phi_{jb}(r_j)\sum_{t_i}p_i(t_i)
\left[\mathbf 1_{\{t_i=a\}}-p_i(a)\right]
=0.
\]
Thus,
\begin{equation}\label{eq:ut-zero-expectations} \sum_{t_{-i}}p_{-i}(t_{-i})H_i(r_i,t_{-i})=0 = \sum_{t_i}p_i(t_i)H_i(t_i,r_{-i})\end{equation}
for every agent $i$, every report $r_i\in T_i$, and every profile $r_{-i}\in T_{-i}$ of opponents' reports.

Now consider an auxiliary game with no private information. Each agent $i$ simply chooses $r_i\in T_i$ and receives $H_i(r)$ at the resulting profile. 
\begin{claim}\label{eq:ut-auxiliary-marginals}
    Every Nash equilibrium \(\mu=(\mu_i)_i\) of the auxiliary game satisfies \(\mu_i=p_i\) for all \(i\neq i_0\).
\end{claim}
\begin{proof}
Fix a Nash equilibrium $\mu$ of the auxiliary game. By the second equality in \eqref{eq:ut-zero-expectations}, \(\sum_{r_i}p_i(r_i)\sum_{r_{-i}}\left(\prod_{j\neq i}\mu_j(r_j)\right)H_i(r_i,r_{-i})=0\), so agent $i$'s equilibrium payoff is nonnegative. Moreover, \(\sum_i\sum_r\left(\prod_j\mu_j(r_j)\right)H_i(r)=\sum_r\left(\prod_j\mu_j(r_j)\right)\sum_iH_i(r)=0\), so every agent's equilibrium payoff is zero. Hence, for every $r_i\in T_i$,
\begin{equation}\label{eq:claimin}
    \sum_{r_{-i}}\left(\prod_{j\neq i}\mu_j(r_j)\right)H_i(r_i,r_{-i}) \leq 0.
\end{equation}
But \(\sum_{r_i}p_i(r_i)\sum_{r_{-i}}\left(\prod_{j\neq i}\mu_j(r_j)\right)H_i(r_i,r_{-i})=0\), and since $p_i(r_i)>0$ for every $r_i$, \eqref{eq:claimin} must hold with equality for every $r_i\in T_i$.

Take reports $a\in T_i\setminus\{t_i^0\}$ and $b\in T_j\setminus\{t_j^0\}$ that were chosen together in the construction above, with $i<j$. The construction adds $\phi_{ia}(r_i)\phi_{jb}(r_j)$ to $H_i(r)$ and subtracts it from $H_j(r)$. Because no report of an agent was chosen twice, this is the only term in $H_i$ containing $\phi_{ia}$ and the only term in $H_j$ containing $\phi_{jb}$. Moreover, \(\phi_{ia}(a)-\phi_{ia}(t_i^0)=1\), while \(\phi_{ic}(a)-\phi_{ic}(t_i^0)=0\) for every $c\in T_i\setminus\{t_i^0,a\}$. Since $a$ and $t_i^0$ both deliver agent $i$ expected payoff zero against $\mu_{-i}$,
\[0 = \sum_{r_{-i}}\left(\prod_{k\neq i}\mu_k(r_k)\right)\left[H_i(a,r_{-i})-H_i(t_i^0,r_{-i})\right] = \sum_{r_j}\mu_j(r_j)\phi_{jb}(r_j) = \mu_j(b)-p_j(b).\]
Hence, $\mu_j(b)=p_j(b)$. Mirroring this logic, $\mu_i(a)=p_i(a)$.

We have deduced the equality between $\mu_i$ and $p_i$ at both reports chosen in each pair. By the choice of $i_0$, every report in $T_i\setminus\{t_i^0\}$ of every agent $i\neq i_0$ was chosen in one of the pairs. Therefore, $\mu_i(a)=p_i(a)$ for every $i\neq i_0$ and every $a\in T_i\setminus\{t_i^0\}$. As $\mu_i$ and $p_i$ are distributions, \(\mu_i(t_i^0)=1-\sum_{a\in T_i\setminus\{t_i^0\}}\mu_i(a)=1-\sum_{a\in T_i\setminus\{t_i^0\}}p_i(a)=p_i(t_i^0)\).
\end{proof}

For $K\geq1$, let $\kappa_{i_0}(K)=K$ and $\kappa_i(K)=K^2$ for $i\neq i_0$, and modify payments to
\begin{equation}\label{eq:ut-modified-payments} \hat\tau_i^K(r)\coloneqq\tau_i(r)-\kappa_i(K)H_i(r). \end{equation}
By \eqref{eq:ut-modified-payments} and the first equality in \eqref{eq:ut-zero-expectations},
\[\sum_{t_{-i}}p_{-i}(t_{-i})\hat\tau_i^K(r_i,t_{-i}) =\sum_{t_{-i}}p_{-i}(t_{-i})\tau_i(r_i,t_{-i}) -\kappa_i(K)\sum_{t_{-i}}p_{-i}(t_{-i})H_i(r_i,t_{-i})=\sum_{t_{-i}}p_{-i}(t_{-i})\tau_i(r_i,t_{-i}),\]
which establishes \eqref{eq:ut-preserved-payments}. It also leaves every interim comparison between reports unchanged when the opponents report truthfully. Hence, truth telling remains a strict BNE for every $K$.

I show that truth is the only equilibrium for all sufficiently large finite $K$. Suppose otherwise. Then there is a sequence $K_m\to\infty$ such that, writing $\hat\tau^m$ for the payment rule in \eqref{eq:ut-modified-payments} at $K=K_m$, $(q,\hat\tau^m)$ has a nontruthful equilibrium $\sigma^m$ for every $m$. Define the \textit{ex ante} probability that agent $i$ reports $r_i$ by \(\mu_i^m(r_i)\coloneqq\sum_{t_i}p_i(t_i)\sigma_i^m(r_i\mid t_i)\) and  $\mu^m\coloneqq(\mu_i^m)_i\in\prod_i\Delta(T_i)$ for the resulting profile of report distributions. Because the agents' types are independent, the probability of a report profile $r$ is
\[\sum_{t\in T}\prod_i p_i(t_i)\sigma_i^m(r_i\mid t_i)=\prod_i\sum_{t_i}p_i(t_i)\sigma_i^m(r_i\mid t_i)=\prod_i\mu_i^m(r_i).\]
If agent $i$ has type $t_i$ and reports $r_i$, his expected net utility under the original mechanism $(q,\tau)$ is \(\sum_{r_{-i}}\left(\prod_{j\ne i}\mu_j^m(r_j)\right) \left[u_i(q(r_i,r_{-i}),t_i)-\tau_i(r_i,r_{-i})\right]\).

Choose $B>0$ such that $|u_i(q(r),t_i)-\tau_i(r)|\leq B$ for every agent, type, and report profile.\footnote{Such a bound exists because the original mechanism is fixed and its type and report sets are finite.} Passing to a subsequence, write $\mu^m\to\mu$. For a profile $\nu=(\nu_j)_j\in\prod_j\Delta(T_j)$ of report distributions, define
\[H_i(\nu)\coloneqq\sum_{r\in T}\left(\prod_j\nu_j(r_j)\right)H_i(r) \quad\text{\&}\quad
H_i(r_i,\nu_{-i})\coloneqq
\sum_{r_{-i}\in T_{-i}}
\left(\prod_{j\neq i}\nu_j(r_j)\right)H_i(r_i,r_{-i}).
\]
Fix an agent $i$ and a report $r_i\in T_i$. Since $\sigma^m$ is a BNE of $(q,\hat\tau^m)$, each type $t_i\in T_i$ receives at least as much expected utility from following $\sigma_i^m(\cdot\mid t_i)$ as from reporting $r_i$, so
\[
\begin{aligned}
&\sum_{s_i,r_{-i}}\sigma_i^m(s_i\mid t_i)
\left(\prod_{j\neq i}\mu_j^m(r_j)\right)
\left[
u_i(q(s_i,r_{-i}),t_i)-\tau_i(s_i,r_{-i})
+\kappa_i(K_m)H_i(s_i,r_{-i})
\right]\\
&\qquad\geq
\sum_{r_{-i}}
\left(\prod_{j\neq i}\mu_j^m(r_j)\right)
\left[
u_i(q(r_i,r_{-i}),t_i)-\tau_i(r_i,r_{-i})
+\kappa_i(K_m)H_i(r_i,r_{-i})
\right].
\end{aligned}
\]
Multiply these inequalities by $p_i(t_i)$ and sum over $t_i$. The terms containing $H_i$ become $\kappa_i(K_m)H_i(\mu^m)$ and $\kappa_i(K_m)H_i(r_i,\mu_{-i}^m)$. By the choice of $B$,
\[
\begin{aligned}
-B
&\leq
\sum_{t_i,s_i,r_{-i}}p_i(t_i)\sigma_i^m(s_i\mid t_i)
\left(\prod_{j\neq i}\mu_j^m(r_j)\right)
\left[u_i(q(s_i,r_{-i}),t_i)-\tau_i(s_i,r_{-i})\right]
\leq B, \quad \text{\&}\\
-B
&\leq
\sum_{t_i,r_{-i}}p_i(t_i)
\left(\prod_{j\neq i}\mu_j^m(r_j)\right)
\left[u_i(q(r_i,r_{-i}),t_i)-\tau_i(r_i,r_{-i})\right]
\leq B.
\end{aligned}
\]
Consequently, \(B+\kappa_i(K_m)H_i(\mu^m) \geq-B+\kappa_i(K_m)H_i(r_i,\mu_{-i}^m)\) and so \(H_i(\mu^m)\geq H_i(r_i,\mu_{-i}^m)-\frac{2B}{\kappa_i(K_m)}\).

As $m\to\infty$, $\kappa_i(K_m)\to\infty$. Both $H_i(\nu)$ and $H_i(r_i,\nu_{-i})$ are continuous in $\nu$, so $\mu^m\to\mu$ implies \(H_i(\mu^m)\to H_i(\mu)\) and \(H_i(r_i,\mu_{-i}^m)\to H_i(r_i,\mu_{-i})\). Taking $m\to\infty$ in \(H_i(\mu^m)\geq H_i(r_i,\mu_{-i}^m)-\frac{2B}{\kappa_i(K_m)}\) yields $H_i(\mu)\geq H_i(r_i,\mu_{-i})$ for every $i$ and $r_i$. Thus, $\mu$ is a Nash equilibrium of the auxiliary game. By \Cref{eq:ut-auxiliary-marginals}, $\mu_i=p_i$ for every $i\neq i_0$. Since $\mu^m\to\mu$, $\mu_i^m\to p_i$ for every $i\neq i_0$.

Let $e_i^m\coloneqq\sum_{t_i}p_i(t_i)[1-\sigma_i^m(t_i\mid t_i)]$ be agent $i$'s probability of misreporting. Define
\[
\begin{aligned}
d_i^m\coloneqq
\sum_{t_i,r_i,r_{-i}}
&p_i(t_i)\sigma_i^m(r_i\mid t_i)
\left(\prod_{j\neq i}\mu_j^m(r_j)\right)\\
&\times
\left[
u_i(q(t_i,r_{-i}),t_i)-\tau_i(t_i,r_{-i})
-u_i(q(r_i,r_{-i}),t_i)+\tau_i(r_i,r_{-i})
\right].
\end{aligned}
\]
Each coefficient \(p_i(t_i)\sigma_i^m(r_i\mid t_i)\left(\prod_{j\neq i}\mu_j^m(r_j)\right)\) is nonnegative and sums to one over $t_i$, $r_i$, and $r_{-i}$. Each bracketed expression lies between $-2B$ and $2B$, so $|d_i^m|\leq2B$.

For every agent $i$ with more than one type, strict IC is
\[
\sum_{r_{-i}}
\left(\prod_{j\neq i}p_j(r_j)\right)
\left[
u_i(q(t_i,r_{-i}),t_i)-\tau_i(t_i,r_{-i})
-u_i(q(r_i,r_{-i}),t_i)+\tau_i(r_i,r_{-i})
\right]>0
\]
for every $t_i\in T_i$ and every $r_i\neq t_i$. These expressions are continuous in the opponents' report distributions. Since there are finitely many agents, types, and reports, there are $c>0$ and, for each agent $i$ with more than one type, a neighborhood $N_i$ of $(p_j)_{j\neq i}$ such that
\[
(\nu_j)_{j\neq i}\in N_i
\quad\Longrightarrow\quad
\sum_{r_{-i}}
\left(\prod_{j\neq i}\nu_j(r_j)\right)
\left[
u_i(q(t_i,r_{-i}),t_i)-\tau_i(t_i,r_{-i})
-u_i(q(r_i,r_{-i}),t_i)+\tau_i(r_i,r_{-i})
\right]\geq c
\]
for every $t_i\in T_i$ and every $r_i\neq t_i$. Therefore,
\begin{equation}\label{eq:ut-strict-gain}
(\mu_j^m)_{j\neq i}\in N_i
\quad\Longrightarrow\quad
d_i^m\geq
c\sum_{t_i}p_i(t_i)\sum_{r_i\neq t_i}\sigma_i^m(r_i\mid t_i)
=c e_i^m.
\end{equation}
If $T_i=\{t_i\}$, then $\sigma_i^m(t_i\mid t_i)=1$, so $e_i^m=d_i^m=0$. For this agent, set $N_i=\prod_{j\neq i}\Delta(T_j)$.

If every type of agent $i$ reports truthfully while the other agents use $\sigma_{-i}^m$, then agent $i$'s type is distributed according to $p_i$, while the opponents' reports are distributed according to $\prod_{j\neq i}\mu_j^m$. Therefore, \(\sum_{r_{-i}}\left(\prod_{j\neq i}\mu_j^m(r_j)\right)\sum_{t_i}p_i(t_i)H_i(t_i,r_{-i})=0\) by the second equality in \eqref{eq:ut-zero-expectations}. Hence, the change in agent $i$'s \textit{ex ante} utility under $(q,\hat\tau^m)$ from replacing $\sigma_i^m$ with truthful reporting is \(d_i^m-\kappa_i(K_m)H_i(\mu^m)\), which is weakly negative since $\sigma_i^m$ is a best response to $\sigma_{-i}^m$. Accordingly, \(d_i^m\leq\kappa_i(K_m)H_i(\mu^m)\), and
\begin{equation}\label{eq:ut-weighted-gains}
\sum_i\frac{d_i^m}{\kappa_i(K_m)} \leq\sum_iH_i(\mu^m) =\sum_{r\in T}\left(\prod_j\mu_j^m(r_j)\right)\sum_iH_i(r) =0, \quad\text{so}\quad K_m d_{i_0}^m+\sum_{i\neq i_0}d_i^m\leq0.
\end{equation}

Because $\mu_j^m\to p_j$ for every $j\neq i_0$, $(\mu_j^m)_{j\neq i_0}\in N_{i_0}$ for all sufficiently large $m$. For those $m$,
\[ 0\leq c e_{i_0}^m\leq d_{i_0}^m \leq-\frac1{K_m}\sum_{i\neq i_0}d_i^m \leq\frac1{K_m}\sum_{i\neq i_0}|d_i^m| \leq\frac{2B(n-1)}{K_m}\rightarrow0.\]
Truthful reports have distribution $p_{i_0}$, so $\|\mu_{i_0}^m-p_{i_0}\|_1\leq2e_{i_0}^m\to0$. Thus, all agents' report distributions converge to their truthful distributions.

We have shown that $\mu_j^m\to p_j$ for every agent $j$. Hence, for all sufficiently large $m$, $(\mu_j^m)_{j\neq i}\in N_i$ for every agent $i$, and \(d_i^m\geq c e_i^m\geq0\) for every \(i\). At the same time, \eqref{eq:ut-weighted-gains} implies \(\sum_i\frac{d_i^m}{\kappa_i(K_m)}\leq0\). Every term in this sum is weakly positive, so $d_i^m=0$ for every $i$. Since $0=d_i^m\geq c e_i^m$ and $c>0$, $e_i^m=0$ for every $i$. Finally, \(0=e_i^m=\sum_{t_i}p_i(t_i)[1-\sigma_i^m(t_i\mid t_i)]\). Every term in this sum is weakly positive, and $p_i(t_i)>0$ for every type. Hence, $\sigma_i^m(t_i\mid t_i)=1$ for every $i$ and $t_i$, contradicting the choice of $\sigma^m$. Therefore, for sufficiently large finite $K$, the payment rule in \eqref{eq:ut-modified-payments} makes truth telling the unique BNE. Taking any such $K$ completes the proof.\end{proof}

\begin{proof}[Proof of \Cref{prop:ut-revelation}]
Fix a finite mechanism $M=((A_i)_i,q,\tau)$ and an equilibrium $\sigma$ satisfying IR. Directify this equilibrium by setting
\[q^0(x\mid r)\coloneqq\sum_{a\in A}q(x\mid a)\prod_i\sigma_i(a_i\mid r_i),\qquad \text{\&} \qquad \tau_i^0(r)\coloneqq\sum_{a\in A}\tau_i(a)\prod_j\sigma_j(a_j\mid r_j).\]
If agents report truthfully, this direct mechanism $D^0=(q^0,\tau^0)$ reproduces the conditional allocation probabilities and every type's interim expected payment. A report $r_i$ by type $t_i$ induces exactly the allocation lottery and expected payment from using $\sigma_i(\cdot\mid r_i)$ in the original mechanism against $\sigma_{-i}$. That mixed message was available to type $t_i$ in $M$, so the equilibrium inequalities make truth telling optimal in $D^0$. As a result, \(U_i(t_i;D^0,\sigma^T)=U_i(t_i;M,\sigma)\) for every $i$ and $t_i$, and \(\Pi(D^0,\sigma^T)=\Pi(M,\sigma)\).

I next construct a direct mechanism that makes every truth-telling inequality strict. For each agent and type, define
\[w_i(t_i)_x\coloneqq u_i(x,t_i)-\frac1{|X|}\sum_{y\in X}u_i(y,t_i) \qquad(x\in X).\]
By construction, \(\sum_{x\in X}w_i(t_i)_x=0\) for every $i$ and $t_i$. By assumption, $w_i(t_i)\neq w_i(s_i)$ whenever $t_i\neq s_i$. Choose a lottery $q^\circ$ with full support on $X$. Since there are finitely many agents and types, there is an $\eta>0$ such that $q^\circ+\eta w_i(r_i)\in\Delta(X)$ for every $i,r_i$. Define a direct mechanism $L=(q^L,\tau^L)$ by (with the norm Euclidean):
\[q^L(r)\coloneqq\frac1n\sum_i\left(q^\circ+\eta w_i(r_i)\right),\qquad \text{\&} \qquad \tau_i^L(r)\coloneqq\frac{\eta}{2n}\|w_i(r_i)\|^2.\]

For an arbitrary vector of opponents' reports, type $t_i$'s net benefit from truth over a different report $r_i$ is
\[\begin{aligned}
&u_i(q^L(t_i,r_{-i}),t_i)-\tau_i^L(t_i,r_{-i})
-u_i(q^L(r_i,r_{-i}),t_i)+\tau_i^L(r_i,r_{-i})\\
&\qquad=\frac{\eta}{n}\langle w_i(t_i),w_i(t_i)-w_i(r_i)\rangle
-\frac{\eta}{2n}\left(\|w_i(t_i)\|^2-\|w_i(r_i)\|^2\right)\\
&\qquad=\frac{\eta}{2n}\|w_i(t_i)-w_i(r_i)\|^2>0,
\end{aligned}\]
where the first equality uses \(\sum_{x\in X}\left[w_i(t_i)_x-w_i(r_i)_x\right]=0\). Accordingly, truth telling is strictly dominant under $L$.

Write $U_i^L(t_i)\coloneqq U_i(t_i;L,\sigma^T)$ and let $b_i\coloneqq\max_{t_i\in T_i}\max\{0,o_i(t_i)-U_i^L(t_i)\}$. For $\delta\in(0,1)$, define \(q^\delta\coloneqq(1-\delta)q^0+\delta q^L\) and \(\tau_i^\delta\coloneqq(1-\delta)\tau_i^0+\delta\tau_i^L-\delta b_i-\delta^2\). Since $D^0$ is IC, $L$ makes every nontrivial reporting inequality strict, and each $-\delta b_i-\delta^2$ does not depend on agent $i$'s report, $D^\delta=(q^\delta,\tau^\delta)$ is strictly BIC.

Apply \Cref{lem:ut-transfer-modification} to $D^\delta$, obtaining $\hat D^\delta=(q^\delta,\hat\tau^\delta)$. Truth telling is its unique BNE. \eqref{eq:ut-preserved-payments} preserves the truthful interim payments and utilities of $D^\delta$, so this equilibrium remains interim IR. For each type,
\[\sum_{t_{-i}}p_{-i}(t_{-i})\hat\tau_i^\delta(t_i,t_{-i})=(1-\delta)\bar\tau_i^\sigma(t_i)+\delta\sum_{t_{-i}}p_{-i}(t_{-i})\tau_i^L(t_i,t_{-i})-\delta b_i-\delta^2.\]
As $\delta\downarrow0$, these payments converge to $\bar\tau_i^\sigma(t_i)$, while $q^\delta(x\mid t)\to q^0(x\mid t)=q^\sigma(x\mid t)$ for every $x,t$. Consequently, $\Pi(\hat D^\delta,\sigma^T)\to\Pi(M,\sigma)$. Choose $\delta>0$ sufficiently small that \(\left|\Pi(\widehat D^\delta,\sigma^T)-\Pi(M,\sigma)\right|<\varepsilon\), and set $D_\varepsilon\coloneqq\widehat D^\delta$.\end{proof}

\bibliography{revprin}

\newpage

\section{Supplement: Private Values}\label[appendix]{sec:private-values}

I now prove the assertions in \Cref{thm:ordinary} concerning private values and independent types. I modify the construction in \secref{sec:construction} so that an agent's utility depends only on his own type and the chosen outcome. The types remain independent, but I replace the uniform marginal distribution with one that assigns most of its probability to type zero.

For every payoff in the prescribed finite set $F$, the modified direct mechanism still has a strict equilibrium that gives the principal that payoff. And although it may have additional equilibrium payoffs, by concentrating enough probability on type zero, I can make every additional payoff arbitrarily close to $F$.

\begin{lemma}\label{lem:private-values-construction}
Let $F=\{a_0,a_1,\ldots,a_k\}$, with $b\coloneqq a_0>a_1>\cdots>a_k$ and $k\geq1$, and fix $\eta>0$. There is an environment $E$ with two agents, private values, and independent types on $\{0,\ldots,k\}$, a truthful direct mechanism $D^0$, a mechanism $M^+$ with one additional message for each agent, and a $\delta>0$, satisfying:
\begin{enumerate}[nosep,noitemsep]
\item At every type profile, the principal's payoff is at most $b$, and exactly one outcome gives her $b$. These maximizing outcomes differ across type profiles.
\item $F\subseteq S(D^0)$, $d_H(S(D^0),F)<\eta/2$, and every member of $F$ is attained at a strict pure equilibrium of $D^0$.
\item $M^+$ has a unique BNE, with $S(M^+)=\{b\}$.
\item Every direct mechanism $D$ satisfies
\begin{equation}\label{eq:private-values-near-maximum} \max S(D)>b-\delta \quad \implies \quad d_H(S(D),F)<\eta. \end{equation}
\end{enumerate}
\end{lemma}
\begin{proof} Let $N\coloneqq k+1$, let $T_1=T_2=\{0,\ldots,k\}$,
and take $X\coloneqq(T_1\cup\{C\})\times(T_2\cup\{C\})$, where $C$ is a symbol outside the sets of types. For now, fix $0<\varepsilon<1/(2k^2+3)$. Let the agents' types be independent, with common distribution \(p_\varepsilon(0)=1-\varepsilon\) and \(p_\varepsilon(j)=\frac{\varepsilon}{k}\) for \(j=1,\ldots,k\). I will choose $\varepsilon$ after establishing the properties of this family of environments.

For $x_i\in T_i$, let $j_i$ be the unique integer in $\{0,\ldots,k\}$ congruent to $x_i-t_i$ modulo $N$. If $x_1\in T_1$ and $x_2\in T_2$, give agent $i$ utility \(u_i(x,t_i)=-\left(j_i-x_{-i}\right)^2\). If $x_1=C$ or $x_2=C$, set \[u_i(x,t_i)= \begin{cases} -\tfrac12,&x_i=C,\quad x_{-i}\in\{0,C\},\\ 1,&x_i=C,\quad x_{-i}\in\{1,\ldots,k\},\\ 0,&x_i=t_i,\quad x_{-i}=C,\\ -1,&x_i\in T_i\setminus\{t_i\},\quad x_{-i}=C. \end{cases}\]

Retain the principal's payoffs from \Cref{lem:construction}. If $x_1\in T_1$ and $x_2\in T_2$, set $v(x,t)=W_{j_1j_2}$, where the diagonal and adjacent entries of $W$ are given by \eqref{eq:principal-matrix}, and every other off-diagonal entry is $a_k$. If $x_1=C$ or $x_2=C$, set $v(x,t)=a_k$. Thus, $v(x,t)\leq b$, with equality precisely when $x=t$. This proves part~\textnormal{(i)} for every $\varepsilon$ in the specified range.

Let $D^0$ choose $x=(r_1,r_2)$ with probability 1 after reports $(r_1,r_2)\in T$. In $M^+$, each agent $i$ can send a message in $T_i\cup\{C\}$, and messages $(a_1,a_2)$ lead to $x=(a_1,a_2)$ with probability 1. Write $S_\varepsilon\coloneqq S(D^0)$ when the marginal distribution of types is $p_\varepsilon$.

\begin{claim}\label{claim:private-values-shifts}
Every member of $F$ is attained at a strict pure equilibrium of $D^0$. In particular, $F\subseteq S_\varepsilon$, and $D^0$ is truthful.
\end{claim}
\begin{proof}
Fix $j\in\{0,\ldots,k\}$ and have every type of each agent report $(t_i+j)\bmod N$. Fix an agent $i$. The other agent reports $j$ with probability $1-\varepsilon$. Let $\mu_j\coloneqq\E[x_{-i}]$ be that agent's expected report. Then $|\mu_j-j|\leq\varepsilon k<1/2$. Each type $t_i$ of agent $i$ can choose any difference $d\in\{0,\ldots,k\}$ by reporting $(t_i+d)\bmod N$, yielding expected utility $-d^2+2d\mu_j-\E[x_{-i}^2]$. Because $|\mu_j-j|<1/2$, this expected utility is uniquely maximized at $d=j$. Accordingly, the specified reports form a strict pure equilibrium. Both differences equal $j$ at every type profile, so the principal receives $W_{jj}=a_j$. Taking $j=0$ produces the truthful equilibrium.
\end{proof}

\begin{claim}\label{claim:private-values-unique}
$M^+$ has a unique BNE, in which every type reports truthfully.
\end{claim}
\begin{proof}
Do iterated deletion of strictly dominated strategies: every false report is strictly dominated by \(C\). Excluding them, every type strictly prefers truth to $C$.
\end{proof}

\begin{claim}\label{claim:private-values-convergence}
$d_H(S_\varepsilon,F)\to0$ as $\varepsilon\downarrow0$.
\end{claim}
\begin{proof}
Take any sequence $\varepsilon_m\downarrow0$ and, for each $m$, any equilibrium $\sigma^m$ of $D^0$ under the corresponding prior. The space of strategy profiles is a fixed finite product of simplices, so a subsequence converges to a profile $\sigma^*$. Conditional on either agent having type zero, the other agent's type is zero with probability approaching one. Accordingly, passing to the limit in the best-response inequalities for the two type-zero agents reveals that $\left(\sigma_1^*(\cdot\mid0),\sigma_2^*(\cdot\mid0)\right)$ is an equilibrium of the complete-information game in which each agent chooses a report in $\{0,\ldots,k\}$ and receives $-(r_i-r_{-i})^2$.

The nearest-integer calculation in the proof of \Cref{claim:direct-equilibria} enumerates the equilibria of this game. Either both agents choose the same pure report $j$, or both mix equally over the same adjacent reports $j,j+1$. The principal receives $a_j$ in the first case and $a_{j+1}$ in the second, by \eqref{eq:mixed-principal-payoff}. Because the probability of the true profile $(0,0)$ approaches one and the principal's payoffs are bounded independently of $\varepsilon$, the principal's expected payoff converges to that same member of $F$.

As a consequence, \(\max_{s\in S_\varepsilon}\dist(s,F)\rightarrow 0\).\footnote{If this failed, there would be a sequence of equilibrium payoffs staying a positive distance from $F$. Choosing equilibria attaining those payoffs and taking a convergent subsequence would contradict the limit just established.} By \Cref{claim:private-values-shifts}, every member of $F$ belongs to $S_\varepsilon$, so $\max_{a\in F}\dist(a,S_\varepsilon)=0$.
\end{proof}

I now choose one $\varepsilon>0$ in the specified range for which $d_H(S_\varepsilon,F)<\eta/2$, and fix the resulting environment $E$. Together with \Cref{claim:private-values-shifts}, this establishes part~\textnormal{(ii)}.

\begin{claim}\label{claim:private-values-uniform}
There is a $\delta>0$ for which \eqref{eq:private-values-near-maximum} holds in this fixed environment.
\end{claim}
\begin{proof}
Suppose otherwise. There are direct mechanisms $D_m$ with
\[\max S(D_m)>b-\frac1m, \qquad \text{\&} \qquad d_H(S(D_m),F)\geq\eta.\]
Choose an equilibrium $\sigma_m$ attaining the maximum in each mechanism. Compactness of the spaces of direct rules and strategies yields, along a subsequence, $D_m\to D^*$ and $\sigma_m\to\sigma^*$. Passing to the limit in \eqref{eq:bne} and \eqref{eq:principal-expectation} reveals that $\sigma^*$ is an equilibrium of $D^*$ giving the principal $b$. The argument in the proof of \Cref{claim:all-direct-rules} applies here and so $D^*$ is a relabeling of $D^0$.

Every $a\in F$ is attained at a strict pure equilibrium of $D^*$. The finitely many strict best-response inequalities remain satisfied when the direct rule is sufficiently close to $D^*$. Thus, for all sufficiently large $m$, the same pure profile is an equilibrium of $D_m$, with principal payoff converging to $a$. Since $F$ is finite, \(\max_{a\in F}\dist(a,S(D_m))\rightarrow 0\).

Because $d_H(S(D_m),F)\geq\eta$, for all sufficiently large $m$ there must be a payoff $s_m\in S(D_m)$ with $\dist(s_m,F)\geq\eta$. Choose an equilibrium attaining each $s_m$ and take a further subsequence of these strategies that converges. The equilibrium inequalities hold at the limit, so the payoffs converge to some $s^*\in S(D^*)=S_\varepsilon$. But $\dist(s^*,F)<\eta/2$ by the choice of $\varepsilon$, contradicting $\dist(s_m,F)\geq\eta$. This proves part~\textnormal{(iv)}.
\end{proof}

All four conclusions hold in the fixed environment $E$.
\end{proof}

\subsection{The characterization with private values}

\begin{proof}[Proof of the private-values assertions in \Cref{thm:ordinary}]
The proof in \secref{subsec:ordinary-proof} that \conditionref{cond:ordinary-iii} implies \conditionref{cond:ordinary-ii} applies to every finite environment, including those with private values and independent types. \conditionref{cond:ordinary-ii} implies \conditionref{cond:ordinary-i} also remains valid on this restricted class.

For necessity, suppose $\V(S)<\max S$ for some compact set $S$. The finite approximations in \secref{subsec:ordinary-proof} give a finite set $F$ with $\V(F)<b\coloneqq\max F$. Let $\gamma\coloneqq(b-\V(F))/2>0$ and choose $\eta>0$ so that \eqref{eq:continuity-at-finite-set} holds. Apply \Cref{lem:private-values-construction} to this $F$ and $\eta$, obtaining an environment $E$, a mechanism $M^+$, and a $\delta>0$.

If a direct mechanism has $\max S(D)\leq b-\delta$, \eqref{eq:internality} implies $\V(S(D))\leq b-\delta$. Otherwise, \eqref{eq:private-values-near-maximum} and \eqref{eq:continuity-at-finite-set} imply $\V(S(D))<b-\gamma$. Consequently, \(\sup_{D\in\D(E)}\V(S(D)) \leq b-\min\{\delta,\gamma\}<b\). The principal's payoff $v(x,t)$ is at most $b$, and $M^+$ has a unique equilibrium giving her $b$. Hence, $V(E)=\V(S(M^+))=b$, establishing \eqref{eq:ordinary-strict-gap} in an environment with private values and independent types. In particular, \conditionref{cond:ordinary-i} fails even when required only on that class of environments.
\end{proof}

\end{document}